\documentclass[reprint, 10pt, floatfix,aps,prd,twocolumn,a4paper,superscriptaddress,nofootinbib,notitlepage]{revtex4-1}
\usepackage{inputenc}
\usepackage[T1]{fontenc} 
\usepackage{paralist}
\usepackage{hyperref}
\usepackage{amsopn,amsthm,dsfont,amssymb}
\usepackage{mathrsfs}
\usepackage{cancel} 
\usepackage{comment} 
\usepackage{color} 
\usepackage{soul} 
\usepackage[normalem]{ulem}
\usepackage{cancel} 
\usepackage{hyperref} 
\usepackage{soul,xcolor}

\hypersetup{
    colorlinks=true,
    linkcolor=blue,
    filecolor=magenta,      
    urlcolor=cyan,
}

\AtBeginDocument{%
    \newwrite\bibnotes
    \def\bibnotesext{Notes.bib}
    \immediate\openout\bibnotes=\jobname\bibnotesext
    \immediate\write\bibnotes{@CONTROL{REVTEX41Control}}
    \immediate\write\bibnotes{@CONTROL{%
    apsrev41Control,author="08",editor="1",pages="1",title="0",year="1"}}
     \if@filesw
     \immediate\write\@auxout{\string\citation{apsrev41Control}}%
    \fi
}%

\newcommand{\ket}[1]{|#1\rangle}
\newcommand{\bra}[1]{\langle #1|}

\usepackage{mathtools}
\usepackage{adjustbox}
\usepackage[caption=false]{subfig}

\begin{document}
\title{Testing spontaneous wavefunction collapse with quantum electromechanics}

\author{Germain Tobar}
\affiliation{Australian Research Council Centre for Engineered Quantum Systems, School of Mathematics and Physics,\\The University of Queensland, St.\ Lucia, QLD 4072, Australia}
\affiliation{Department of Applied Mathematics and Theoretical Physics, Centre for Mathematical Sciences, Wilberforce Road, Cambridge, CB3 OWA, UK}

\author{Stefan Forstner}
\affiliation{ICFO-Institut de Ciencies Fotoniques, The Barcelona Institute of Science and Technology, Castelldefels (Barcelona) 08860, Spain
}

\author{Arkady Fedorov}
\affiliation{Australian Research Council Centre for Engineered Quantum Systems, School of Mathematics and Physics,\\The University of Queensland, St.\ Lucia, QLD 4072, Australia}

\author{Warwick P. Bowen}
\affiliation{Australian Research Council Centre for Engineered Quantum Systems, School of Mathematics and Physics,\\The University of Queensland, St.\ Lucia, QLD 4072, Australia}

\begin{abstract}
Theories of spontaneous wavefunction collapse offer an explanation of the possible breakdown of quantum mechanics for macroscopic systems. However, the challenge of resolving predicted collapse signatures above background noise has precluded conclusive tests. Here, we propose to overcome this challenge using a superconducting qubit to precisely readout the collapse-induced heating of a mechanical resonator. We show that the ability to strongly couple the qubit to the resonator can enable both fast measurements and initialization of the qubit close to its ground state. Combined this greatly suppresses the influence of quasiparticle heating of the qubit, which we predict to be the dominant noise source. We find that bulk acoustic wave resonances can amplify the collapse induced heating due to their ultra-low dissipation. Together, this could enable a conclusive test of collapse models.

\end{abstract}

\maketitle

Despite much recent experimental progress \cite{Fein2019, sqreview, huang2020superconducting}, there is still no widely accepted explanation as to how classical realism emerges for macroscopic systems. Collapse models provide the only experimentally testable solution~\cite{Bassi2013}, postulating that objective collapse of the wavefunction occurs for sufficiently large quantum systems.

Thought experiments provide lower bounds on the collapse rate at a fixed correlation length~\cite{Adler2007, Bassi2010}. Upper bounds are set through experimental tests that search for deviations from quantum theory. These tests fall into two categories: interferometric tests, which aim to produce macroscopic quantum superpositions in order to probe the breakdown of the quantum superposition principle~\cite{Romero-IsartO2011Lqsa, Bassi2013, EibenbergerSandra2013Miop,BatemanJames2014Nioa, KovachyT2015Qsat,BjornPRA, FeinYaakovY.2019Qsom, Kanari-NaishLydiaA2021Ctdd}; and non-interferometric tests, which aim to observe spontaneous heating as an unavoidable side effect of the collapse process \cite{Nimmrichter2014, CarlessoMatteo2022Psaf}. Given the challenge of isolating macroscopic quantum superpositions from environmental decoherence, the strictest tests have so far been performed non-interferometrically using  mechanical resonators~\cite{Vin2016, Helou17, Vin2017, Vin2020}. However, the inability to resolve collapse-induced heating above thermal noise has prevented any conclusive test of CSL~\cite{Adler2007, Bassi2010, Forstner2020}. While optomechanical systems operating at low temperatures and high frequencies have been proposed as a means to circumvent thermal noise, currently available optomechanical coupling strengths are too low for CSL heating to be resolved~\cite{Forstner2020}.

In this work, we propose the use of electromechanical systems for non-interferometric tests of CSL. In our approach, the CSL heating of a mechanical resonator is readout by swapping its state onto a superconducting qubit and measuring the qubit~\cite{Chu2017, Chu2018, Chu2021, wollack2022quantum}. We find that quasiparticle heating of the qubit~\cite{Wenner2013, Jin2015, Serniak2018}  is likely to be the dominant noise channel. To overcome this, we propose a pulsed scheme, where the signal is amplified by allowing CSL heating to build up in the resonator prior to the swap operation. We show that a sequence of measurements on an array of mechanical resonances can be employed to further suppress the quasiparticle noise, with the swap operation for each resonator cooling the qubit to close to its ground state for the next resonator. 

As a specific example implementation, we consider a flux qubit coupled to an array of internal bulk acoustic wave (BAW) resonances~\cite{Chu2017, Chu2018, Chu2021}. With this system, we predict that the qubit noise can be suppressed  by three orders of magnitude relative to the CSL signal. Moreover, we find that the ultra-low dissipation achievable in BAW resonators significantly amplifies the expected CSL occupancy. The combination of noise suppression and signal enhancement is sufficient to close the gap between the measured upper and calculated lower bounds on the collapse rate and therefore conclusively test the CSL model. Furthermore, the access to natural arrays of mechanical resonators in a BAW resonator allows this to be achieved in a matter days, rather than the months or years of most previous proposals~\cite{Forstner2020}.

\begin{figure}[b] 
\begin{center}
\includegraphics[scale  = 0.42]{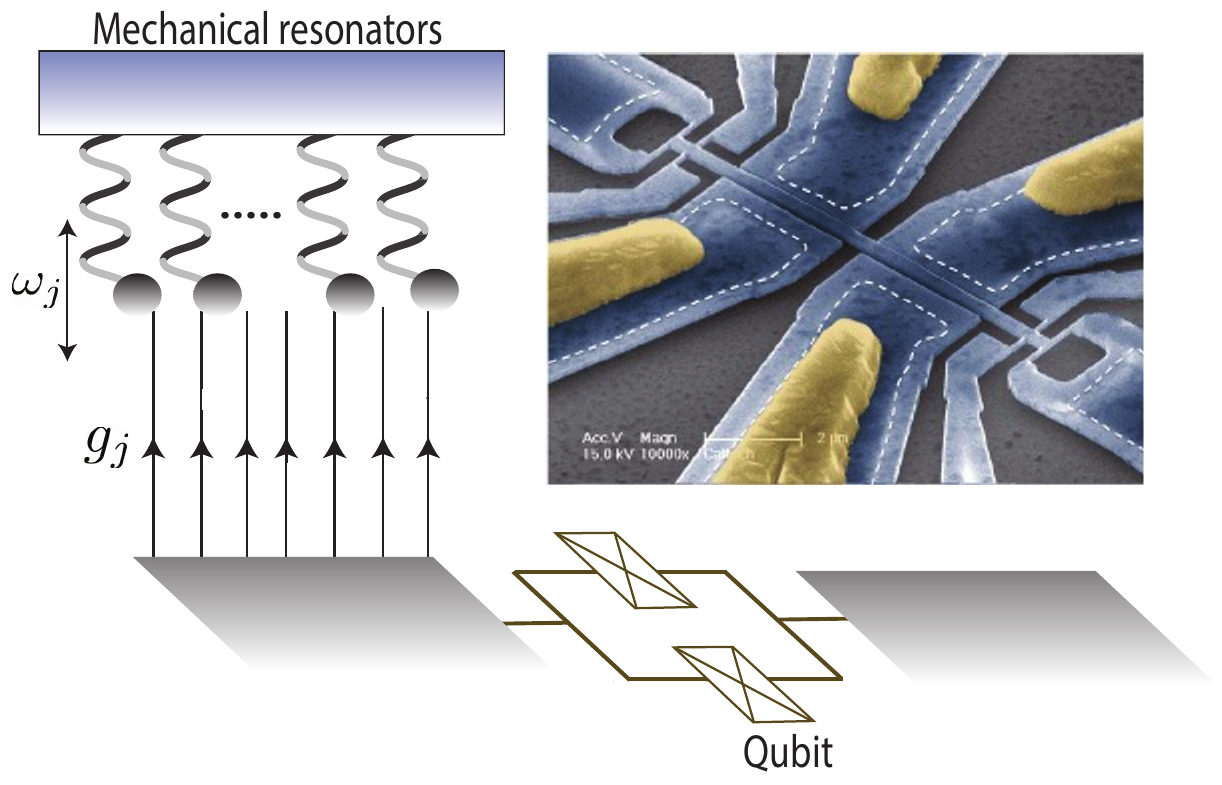}
    \caption{\label{conceptscheme} Scheme to test  CSL-induced heating with an array of qubit-coupled  mechanical resonators. $g_j$: on-resonance coupling rate between the qubit and mechanical resonator $j$ with frequency $\omega_j$.
    }
\end{center}
\end{figure}

\textit{Scheme.} Our scheme utilises Jaynes-Cummings coupling to swap excitations between a mechanical resonator and a superconducting qubit (Fig.~\ref{conceptscheme}). This serves the dual purposes of allowing the qubit to be used to detect the collapse-induced excitations of the mechanical resonator, and cooling it towards the typically lower temperature of the mechanical resonator. To increase the collapse signal relative to qubit noise, we consider a pulsed scheme in which collapse-induced excitations are allowed to build up in the resonator while it is off-resonance with the qubit. The qubit and the mechanical resonator are decoupled until the population of the mechanical resonator reaches equilibrium. The qubit is then tuned on-resonance with the first mechanical resonator, for instance, by flux tuning its frequency \cite{Paw09}. Maintaining resonance for only the duration of a single swap operation transfers qubit excitation that would otherwise be a significant source of noise into the resonator while minimising the time over which qubit-related decoherence can occur. After this swap process, the state of the qubit is measured via a dispersive quantum non-demolition (QND) measurement. 

Using an array of frequency non-degenerate resonators, our scheme can be repeated in rapid succession, cooling the qubit beyond what can be achieved with a single swap operation and multiplying the rate of phonons transfered to the qubit by the number of mechanical resonators. Ultimately, a given CSL collapse rate can be ruled out if it predicts a higher CSL phonon flux than is observed for the relevant correlation length.



\label{Sec:Exp}

While our protocol could be applied using a variety of electromechanical systems \cite{Chu2017, Chu2018,BeraTanmoy2021Lfci}, we consider electromechanical coupling between resonant bulk acoustic waves (BAW) and a superconducting qubit. We choose this example because strong Jaynes-Cummings coupling has been demonstrated in such a system using a piezoelectric interaction~\cite{Chu2017, Chu2018}, BAW resonances can easily reach frequencies at which thermal noise is frozen out at cryogenic temperatures, the high number of mechanical modes within the tunable frequency range of the qubit provides a natural array of mechanical resonators, and because bulk acoustic wave resonances also offer the potential for very high collapse induced heating rates at relevant collapse correlations lengths $r_c$. As we will show in what follows, this arises due to the close matching of their wavelength (at typical GHz frequencies) to the expected collapes correlation length, and to their ultra-low acoustic dissipation.

As a specific example, we consider a qubit-coupled mechanical resonator based on Ref.~\cite{Chu2017}. We proceed by calculating the specifications for a Silicon Carbide (SiC) BAW resonator, due to their ultra-high Q factors and suitability for interfacing with qubits \cite{Gokhale2020}. We choose a resonator thickness of $30 \; \mathrm{\mu m}$, for which the BAW mode with longitudinal and transverse mode numbers $l = 40$ and $m = 0$, respectively, has a frequency $\omega/2\pi = 6.33 \; \mathrm{GHz}$, within the $4-9 \; \mathrm{GHz}$ tunable frequency range of the qubit~\cite{qubitfreqrange}. This corresponds to a phonon wavelength of $\lambda \approx 15 \times 10^{-7} \; \mathrm{m}$, which is sufficiently close to the expected correlation length ($r_c = 10^{-7} \; \mathrm{m}$) to maximise the CSL heating rate of the resonator at this correlation length (see Supplementary Information \cite{Supp} for details).

For a single resonance, the phonon flux predicted by CSL is quantified as $\dot{n}_c = \lambda_c D$ \cite{Forstner2020}, where $D$ is the geometry dependent \textit{CSL cross-section}. Previous calculations of the CSL cross-section have mainly focussed on resonators for which the centre-of-mass motion is dominant~\cite{Carlesso16, Vin2016, Helou17, Vin2017}. A fundamental breathing mode has also been considered, taking the approximation of a uniform linear expansion \cite{Forstner2020}. To allow accurate predictions for BAW resonances, we go beyond these approximations, modeling CSL heating up to quadratic expansions of the acoustic modeshape and for higher order longitudinal modes. We find that at a fixed frequency, lower longitudinal mode numbers imply a higher sensitivity to the collapse-induced heating. This is because the CSL phonon flux decreases with the size of the BAW resonator at a correlation length that matches the BAW wavelength \cite{Supp}. Therefore, thinner BAW resonators are more optimal for tests of CSL. 
 
We find a CSL cross-section at the expected $r_c$ of $D~\approx~5.5~\times~10^5$ (see Supplementary Information \cite{Supp}). For the lower bounds to the collapse rate postulated by Adler~\cite{Adler2007} and Bassi \textit{et al.}~\cite{Bassi2010}, it gives collapse-induced heating rates of $\dot{n}_c \approx 5.5 \times 10^{-3\pm 2} \; s^{-1}$ and $\dot{n}_c \approx 5.5 \times 10^{-5 \pm 2} \; s^{-1}$, respectively. The lowest of these rates equates to the generation of one phonon in under 22 days. Other BAW modes with similar $l$ have comparable CSL cross-sections and heating rates, enabling experiments with arrays of resonances.

\textit{CSL phonon flux.} With the qubit off-resonance, the steady-state flux of phonons leaving the mechanical resonator is $P \gamma_r$ where $\gamma_r$ is the mechanical decay rate and $P$ is the steady-state occupancy of the BAW resonator due to CSL heating. In the absence of other forms of heating, this outwards phonon flux must equal the inwards flux from CSL, so that $P~\approx~\frac{2 \pi \dot{n}_c}{\gamma_r}$. The highest Q-factor measured experimentally in a BAW resonator is $Q = 10^9$ \cite{tobar2021}. Quality factors as high as $Q = 10^7$ have been achieved in SiC BAWs at a temperature of 7 K \cite{Gokhale2020}. We find that $Q = 10^7$ would be insufficient to fully exclude CSL. However, the quality factor of BAWs scales as $T^{-4}$ at low temperatures \cite{Gokhale2020}. Conservatively assuming only a $T^{-1}$ scaling, it was predicted in Ref.~\cite{Gokhale2020} that values as high as $Q=10^{10}$ will be possible at the $\mathrm{mK}$ temperatures required to suppress thermal noise in our system. We choose this higher $Q$ value for our analysis. Using $\dot{n}_c = 5.5 \times 10^{-7} \; \mathrm{s}^{-1}$ for the full exclusion of Bassi \textit{et al.}'s lower bounds and $\gamma_r/2\pi \approx 1 \; \mathrm{Hz}$ ($Q = 10^{10}$ for GHz frequency modes) for the decay rate of a SiC BAW resonator, we find that full exclusion at $r_c = 10^{-7} \; \mathrm{m}$ requires the ability to resolve occupancy's as low as $P \lesssim 5.5 \times 10^{-7}$. 

To determine how efficiently the phonon occupancy can be transferred from the mechanical resonator into the qubit we model the coupling using a phenomenological Linblad master equation in the Supplementary Information \cite{Supp}. We choose parameters consistent with the experiment in Ref.~\cite{Chu2017}, apart from the piezoelectric coupling which can be made an order of magnitude higher through the use of alternative materials (See Supplementary Information \cite{Supp}). One might expect that, with an appropriate choice of interaction time, a swap efficiency approaching 100\% would be possible. However, we find that off-resonance coupling due to residual piezoelectric interaction with adjacent BAW modes (in combination with qubit decay and dephasing) limits the efficiency to $\eta_{\mathrm{swap}} \approx 0.85$. To fully exclude Bassi \textit{et al.}'s lower bound,  all spurious sources of heating in the qubit must therefore be suppressed below the probability $P_{\mathrm{CSL}} = \eta_{\mathrm{swap}} P \approx 4.7 \times 10^{-7}$.

\textit{Noise sources.} The expected noise sources in our protocol are measurement induced heating, state discrimination error, quasiparticle poisoning and the Purcell effect. We find thermal phonons to be negligible when using  BAW modes due to their high frequency (Supplementary Information \cite{Supp}). 

We propose to measure the state of the qubit through coupling to a microwave readout cavity, as is conventional for the measurement of superconducting qubits~\cite{Khezri2016,Boissonneault2009,Slichter2012,KulikovAnatoly2020METo, blais2020circuit}. In the dispersive regime, where the number of photons in the resonator is much less than the critical number $n_{\mathrm{crit}} = {\Delta^2}/{4 g^2}$, with $\Delta$ and $g$ the qubit-cavity detuning and coupling strength respectively, the coupling approximates an ideal quantum non-demolition (QND) measurement \cite{blais2020circuit}. In standard qubit operation, measurement-induced heating due to non-QND effects is considered significant only for photon numbers comparable to the critical number \cite{Khezri2016,Boissonneault2009,Slichter2012}. However, for the ultra-low rate of false positives required to carry out our proposed CSL tests, even minute non-QND effects become significant. The primary non-QND effect is the measurement induced heating that occurs due to higher order terms in $g/\Delta$ in the dispersive approximation \cite{Khezri2016}. Taking these higher order terms into account, we find that for experimental parameters achieved in recent experiments \cite{Walter2017} it is possible to suppress measurement induced heating to below $P_H = 8 \times 10^{-8}$ (see Supplementary Information \cite{Supp}). 


Dispersive readout encodes information about the state of the qubit on 
the phase of the scattered microwave signal. A phase threshold is then used to discriminate whether the qubit was in its excited or ground state. The probability of misidentifying the qubit state is known as the state discrimination error $\epsilon_{\text {sde }}$ \cite{Sank2014}. Unlike quantum computing, where errors in assigning the state of the qubit are generally equally important, for the purpose of distinguishing collapse-induced phonons it is much more important to suppress false positives than false negatives. This motivates the choice of a high discrimination threshold, so that noise alone is unlikely to cause a false positive. However, too high a threshold will also reduce the rate of correctly identifying the qubit in the excited state (true positives), increasing the required duration of the experiment. We find that with an appropriate choice of discrimination threshold it is possible to exponentially suppress the error in false positives to well below $P_{\mathrm{CSL}} \approx 10^{-7}$, while maintaining a reasonable probability of detecting true positives (Supplementary Information \cite{Supp}). For example, the error in false positives can be suppressed to $10^{-7}$ while correctly identifying CSL-heating events with probability $\eta_{\rm disp}=0.1$, so long as the microwave resonator can distinguish the excited state of the qubit from its ground state with a signal-to-noise (SNR) ratio of at least 8.


\setstcolor{red}
The Purcell effect arises from the enhancement of the qubit's spontaneous excitation rate due its coupling to the readout cavity, which is in turn coupled to a transmission line. It has been shown experimentally to produce qubit excitations at a rate of around $\Gamma_{\mathrm{P}}/2\pi \approx 0.5 \; \mathrm{Hz}$. Quasiparticle heating occurs due to hot out-of-equilibrium quasiparticles tunneling across the qubit Josephson junctions into the superconducting island. This causes the qubit to transition to its excited state with a phenomenological excitation rate $\Gamma_{\mathrm{QP}}$ \cite{Serniak2018, KulikovAnatoly2020METo}. Recent measurements have found that $\Gamma_{\mathrm{QP}}/2\pi \approx 300 \; \mathrm{Hz}$ \cite{KulikovAnatoly2020METo} (see Supplementary Information \cite{Supp}), while quasiparticle traps have been shown to suppress quasiparticle excitations by more than an order of magnitude~\cite{HosseinkhaniA2017OCfN, Kalashnikov2019, Mar2020Si}.

Taking $\Gamma_{\mathrm{QP}}/2\pi = 30 \; \mathrm{Hz}$ and $\Gamma_{\mathrm{P}}/2\pi = 0.5 \; \mathrm{Hz}$,
the equilibrium population of the qubit due to quasiparticle and Purcell heating is $P_s \approx \frac{\Gamma_{\mathrm{QP}} + \Gamma_P}{\gamma_q} = 10^{-3}$. This is orders-of-magnitude higher than the steady-state CSL occupancy, suggesting that conclusive tests of CSL may not be possible. However, this is not the case in our proposal due to the build-up of the collapse-induced heating within the mechanical resonator, and because each interaction between the qubit and a mechanical resonance transfers heat out of the qubit, to be dissipated into the mechanical bath.

In order to calculate how effectively our protocol can suppress the Purcell and quasiparticle heating, we solve the qubit-resonator Linblad master equation to simulate a series of swap operations between a qubit with initial state $\rho(0)~=~\frac{\Gamma_{\mathrm{QP}} + \Gamma_P}{\gamma_q} \ket{e}\ket{n = 0}\bra{n = 0}\bra{e}$ and an array of mechanical resonances (see Supplementary Information \cite{Supp}). 
We find that significant cooling is possible, consistent with recent experiments \cite{Chu2017, Chu2018}. In the strong-coupling regime, for which $g_{l,m} > \gamma_q$  (where $g_{l,m}$ is the coupling strength to the $l, m$ mechanical mode), we find that the temperature can be  reduced by a factor of  $\frac{\Gamma_{\mathrm{QP}} + \Gamma_P}{g_{l,m}}$ in the limit of a large number of operations.

\begin{figure}[h]
\begin{center}
\includegraphics[scale  = 0.48]{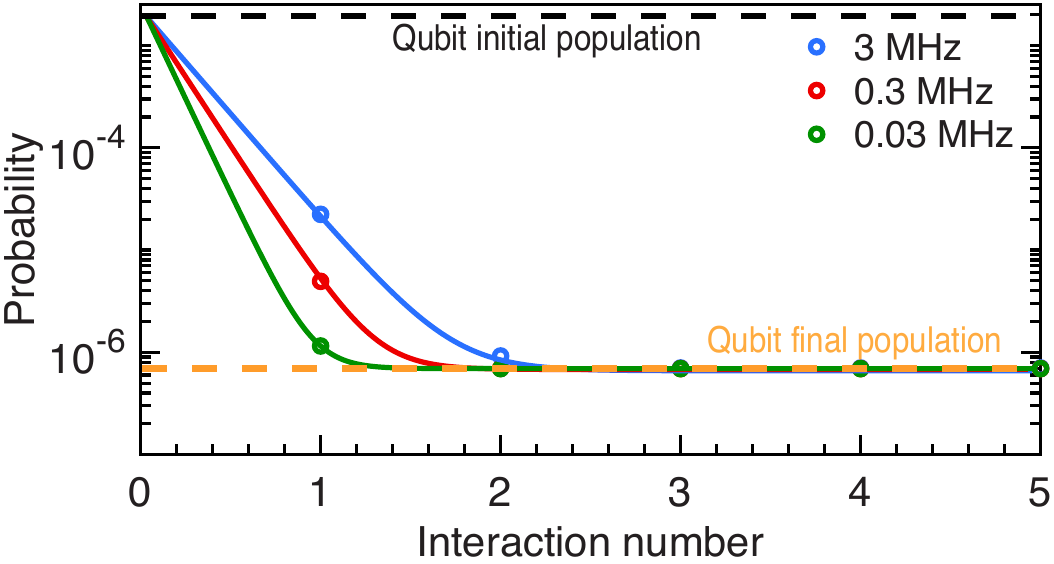}
\caption[Qubit-mechanical resonator coupled dynamics]{\label{effiencyqp}  Reduction in qubit occupancy as it interacts with the series of mechanical resonances for different qubit dephasing rates.  } 
\end{center}
\end{figure}

The qubit in Ref.~\cite{Chu2017} is well within the strong-coupling regime with $g_{l,m}/\gamma_q = 10$. Using their parameters, our previously assumed order of magnitude increase in coupling strength, and $(\Gamma_{\mathrm{QP}} + \Gamma_{\mathrm{P}})/2\pi = 30 \; \mathrm{Hz}$, we find the qubit can be cooled by a factor of $\sim 10^3$, to $P_e \approx 7 \times 10^{-7}$ (see Supplementary Information \cite{Supp}). As shown in Fig.~\ref{effiencyqp}, our simulations suggest that only one or two swap operations will be required to approach this temperature, depending on the qubit dephasing rate. Together with the build-up of CSL heating within the mechanical resonator, this provides a three order of magnitude total suppression of the Purcell and quasiparticle heating relative to the CSL signal.


\begin{table}[ht]
\centering 
\begin{tabular}{c c c c} 
\hline\hline 
Noise type & Scaling & Occupation & $\lambda_{c,\mathrm{min}} \; (s^{-1})$ \\ [0.5ex] 
\hline 
CSL Collapse (signal) & $\eta_{\mathrm{swap}} \lambda_c D$ & $4.7 \times 10^{-7}$ & - \\ 
Thermal & $\exp\left(-\frac{\hbar \omega}{k_B T} \right)$ & $1 \times 10^{-14}$ & $2.1 \times 10^{-20}$  \\
SDE & $ \operatorname{erfc}\left[\frac{\sqrt{\mathrm{SNR}}}{a}\right]$  & $10^{-7}$ & $2.1 \times 10^{-13}$  \\
Measurement  & $\left|\frac{\epsilon g}{\Delta^{2}}\right|^{2}$ & $8 \times 10^{-8}$ & $1.7 \times 10^{-13} $\\
\textbf{Current QP} & $\mathbf{{\Gamma_{Q P}} T_{\mathrm{swap}}}$  & $\mathbf{7 \times 10^{-7}}$ & $\mathbf{1.5 \times 10^{-12}}$  \\
\textbf{Reduced QP} & $\mathbf{{\Gamma_{Q P}} T_{\mathrm{swap}}}$  & $\mathbf{7 \times 10^{-8}}$ & $\mathbf{1.5 \times 10^{-13}}$  \\
All noise (current) & -  & $9 \times 10^{-7}$ & $1.9 \times 10^{-12}$ 
\\
All noise (improved) & -  & $ 2.5 \times 10^{-7}$ & $ 5.3 \times 10^{-13}$ \\[1ex]
\hline 
\end{tabular}
\caption{\label{noisytable} Comparison of noise sources. $\epsilon$ and $T_{\mathrm{swap}}$, dispersive measurement strength and time taken for a swap operation, respectively.}
\end{table}

\textit{Minimum testable collapse rate.} 
Table~\ref{noisytable} displays a quantitative summary of all noise sources in the qubit-resonator system.  Their sum produces a spurious qubit excited state population of $P(\ket{e}) \approx 9 \times 10^{-7}$. Equating this to the CSL-induced excitation probability of $P_{\mathrm{CSL}} \approx  2 \pi \lambda_c D \eta/\gamma_r$, yields a the minimum testable collapse rate of $\lambda_{\mathrm{c,min}} \approx 1.9 \times 10^{-12} \; \mathrm{s}^{-1}$ at the expected correlation length ($r_c = 10^{-7} \; \mathrm{m}$). Our analysis predicts that the dominant noise source is likely to arise from quasiparticle poisoning. While still in their infancy, techniques to suppress this are under active study since it is a key noise source in superconducting quantum computing. Early techniques have already proven to be highly effective~\cite{Court2008}, and variety of recent proposals have been made to achieve further suppression~\cite{HosseinkhaniA2017OCfN, RiwarR.-P2019Eqtw, MannilaE.T2019Dpei, mannila2021superconductor}. Suppression of one order of magnitude beyond the current state of the art would reduce the quasiparticle heating to $\Gamma_{\mathrm{QP}}/2\pi \approx 3 \; \mathrm{Hz}$, corresponding to $\lambda_{c,\mathrm{min,s}} \approx 5.3 \times 10^{-13} \; \mathrm{s}^{-1}$.



\begin{figure}[h]
\begin{center}
\includegraphics[scale  = 1.02]{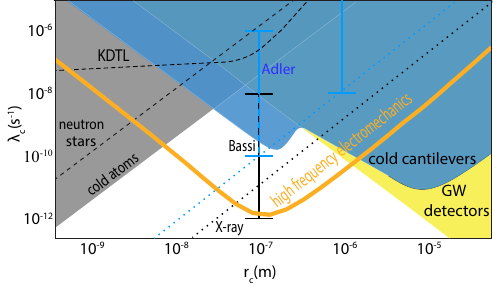}
\caption[Testable parameter space of the CSL model]{\label{boundss2}  CSL exclusion region adapted from Ref.~\cite{Forstner2020}. Excluded upper bounds for both simple CSL and coloured CSL: Cold atoms (grey shaded); gravitational wave detectors (yellow shaded); KDTL interferometry (dashed black); nanocantilevers (blue shaded). Excluded for simple CSL: X-ray (dotted black) and neutron stars (dashed black). Lower bounds: Adler (vertical blue bars and dotted blue line) and Bassi \textit{et al.} (vertical black bar). Orange line: lower limits of the predicted testable parameter space of our protocol.}
\end{center}
\end{figure}

 Fig.~\ref{boundss2}  shows the CSL exclusion region as a function of correlation length. It includes the upper bounds on the collapse rate that our proposal would achieve with existing technology (solid orange line), as well as  the upper bounds from previous experiments and Bassi \textit{et. al}'s and Adler's lower bounds. As can be seen, with the realistic parameters chosen, our proposal has the potential to almost fully close the gap allowing a conclusive test of Bassi \textit{et. al}'s lower bound~\cite{Vin2020,Forstner2020}. With further suppression of quasiparticle poisoning, or improved BAW quality factors, it could extend the upper bounds beyond  Bassi \textit{et. al}'s lower bound~\cite{Vin2020,Forstner2020}.

\textit{Measurement time.} Bassi \textit{et. al.}'s lower bound corresponds to the minimum proposed level of CSL heating that is consistent with the observed classical behaviour of macroscopic systems~\cite{Bassi2010}. For it, the average time until one collapse-induced excitation is observed from a single BAW resonance would be $T_{\mathrm{meas}} = \left(\dot{n}_c \cdot \eta_{\mathrm{swap}} \cdot \eta_{\mathrm{disp}}  \right)^{-1} \approx 5.5 \times 10^{7} \; \mathrm{s}$. Therefore, to conclusively rule out collapse models our experiment would need to be conducted for approximately 247 days. The use of parallel measurements on an array of $N$ mechanical resonances would further reduce the required measurement time by a factor of $N^{-1}$. The BAW resonator we model has around 35 distinct longitudinal modes in the tunable frequency range of the qubit ($4 - 9 \; \mathrm{GHz}$, see Supplementary Information \cite{Supp}) which if used would reduce the experiment time to about 7 days. The access to modes of different frequencies may, further, allow identification of the physical origin of the collapse process and unambiguous differentiation of CSL-induced heating from noise. This could be achieved by measuring the expected frequency dependent CSL-heating rate predicted by coloured models \cite{Adlerr2008}.



\textit{Conclusion.} We  propose a test of spontaneous wavefunction collapse models using qubit-coupled mechanical resonators. The proposal offers the advantage of strong suppression of qubit noise, exquisitely precise readout enabled by the use of a qubit, and access to ultra-high quality bulk acoustic wave resonances that enhance the CSL signal at the expected correlation length. A significant innovation is the use of strong qubit-resonator coupling to cool the qubit. This system may allow for a conclusive test of the CSL model. 

{\bf Acknowledgments.} The authors thank Tyler Jones, Magdalena Zych, Yiwen Chu and Alejandro Gomez for helpful discussions. We thank Matteo Carlesso and Stefan Nimmrichter for feedback on an earlier version of our manuscript. This work was funded by the Australian Research Council Centre of Excellence for
Engineered Quantum Systems (EQUS, project number CE170100009). G.T. acknowledges support from a Cambridge Australia Allen STEM scholarship, jointly funded by Cambridge Australia Scholarships and the Cambridge Trust. S.F acknowledges the Severo Ochoa individual grant (CEX2019-000910-S [MCIN/ AEI/10.13039/501100011033], Fundació Cellex, Fundació Mir-Puig, and Generalitat de Catalunya through CERCA). We acknowledge the traditional owners of the land on which the University of Queensland is situated, the Turrbal and Jagera people.

\newpage

\bibliography{refs}
\bibliographystyle{apsrev4-2}

\appendix

\onecolumngrid




\end{document}


\title{Testing collapse mechanisms with high frequency quantum optomechanics: Supplemental Material}

\author{Germain Tobar$^{1,2}$, Stefan Forstner$^3$, Arkady Federov$^1$, and Warwick P. Bowen$^1$}
\address{$^1$Australian Centre for Engineered Quantum Systems, School of Mathematics and Physics, The University of Queensland, Australia}
\address{$^2$Department of Applied Mathematics and Theoretical Physics, Centre for Mathematical Sciences, Wilberforce Road, Cambridge, CB3 OWA, UK}
\address{$^3$ICFO-Institut de Ciencies Fotoniques, The Barcelona Institute of Science and Technology, Castelldefels (Barcelona) 08860, Spain}
\date{\today} 

\maketitle

\section{Heating rate from spontaneous collapse}\label{CSLappendix}
 
\subsection{CSL-Linblad term for an internal mechanical mode}
Here we calculate the decoherence rate of a bulk acoustic wave resonator by deriving the CSL master equation for the internal mechanical mode of a resonator. We provide the most accurate calculation of the CSL cross-section of an internal mechanical mode of a resonator to date, and derive the generalisation of the CSL cross-section for higher order internal mechanical modes. 

In order to determine the CSL master equation for an internal mechanical mode, we consider the CSL master equation for a body of N masses $m_n$ with position operators $\vec{r}_n$ \cite{Nimmrichter2014}
\begin{ceqn}
\begin{equation} \label{cslmasterch6}
    \mathcal{L}_{\mathrm{csl}} \hat{\rho}=\frac{\lambda_{\mathrm{csl}}}{\pi^{3 / 2} r_{\mathrm{csl}}^{3} \mathrm{amu}^{2}} \int \mathrm{d}^{3} s\left(m(\vec{s}) \hat{\rho} m(\vec{s})-\frac{1}{2}\left\{\hat{\rho}, m^{2}(\vec{s})\right\}\right).
\end{equation}
\end{ceqn}
Here, $\mathcal{L}_{\mathrm{CSL}} $ is the CSL-Lindblad operator, $\hat{\rho}$ is the system's density matrix, $\mathrm{amu}$ is the atomic mass unit and $\lambda_c$, $r_c$ are the CSL collapse rate and noise correlation length respectively. 

The Gaussian-smeared mass density operator for CSL is defined as \cite{Nimmrichter2014, Forstner2020}
\begin{ceqn}
\begin{equation}
    m(\vec{s})=\sum_{n} m_{n} e^{-\frac{(\vec{s}-\vec{r} n)^{2}}{2 r_{c}^2}}=\int_{V} \mathrm{~d}^{3} \vec{r} \rho(\vec{r}) e^{-\frac{(\vec{s}-\vec{r})^{2}}{2 r_{c}^{2}}},
\end{equation}
\end{ceqn}
where $\vec{r}_n$ is the coordinate of the $n$-th mass $m_n$ and $\rho(\vec{r})$ is the mass density. By taking a Fourier transform, the mass density can be expressed as
\begin{ceqn}
\begin{equation} \label{massdensch6}
    m(\vec{s})=\frac{r_{c}^{3}}{(2 \pi)^{3 / 2}} \int \mathrm{d}^{3} \vec{k} e^{-\frac{1}{2} r_{c}^{2} k^{2}} \int_{V} \mathrm{~d}^{3} \vec{r}_{n} \rho\left(\vec{r}_{n}\right) e^{i \vec{k} \cdot \left(\vec{s}-\vec{r}_{n}\right)}.
\end{equation}
\end{ceqn}
The position operator for an infinitesimal volume element of the continuous structure is 
\begin{ceqn}
\begin{equation}
    \vec{r} = \vec{r}_{\mathrm{CoM}} + \vec{r}^{(0)} + \Delta \vec{r}, 
\end{equation}
\end{ceqn}
where $\Delta \vec{r}$ is the displacement of the volume element relative to its equilibrium position, $\vec{r}^{(0)}$ is the equilibrium position of the volume element relative to the centre of mass and $\vec{r}_{\mathrm{CoM}}$ is the centre-of-mass (CoM) operator. For an internal mechanical mode of a resonator there is no oscillation of the system's centre-of-mass requiring meaning that $\vec{r}_{\mathrm{CoM}} = 0$ \cite{Forstner2020}. Therefore, we must examine more closely the effects of $\Delta \vec{r}$. In this work, we will model $\Delta \vec{r}$ as an ideal \textit{breathing mode}. A breathing mode is an expansion of a structure along one or multiple axes with contraction along the other axes \cite{Forstner2020}. The amount of contraction is characterised by its Poisson ratio \cite{Forstner2020}.

\subsubsection{1D breathing mode of a cylinder}
We start by assuming that expansion and contraction of the breathing motion of a cylindrical cavity occurs along a single direction - the vertical direction of phonon polarisation. This assumption holds for materials with low Poisson ratios such as silicon or quartz and simple resonator geometries \cite{Forstner2020}. We consider a cylindrical cavity of height $\lambda/2$ and radius $R$ with periodic expansion and contraction solely in the vertical $x$-direction with no motion in the transverse directions. We define the variable $x_s$ to quantify the displacement at each breathing surface. The position operator of a volume element in a $1-\mathrm{D}$ breathing mode in cylindrical coordinates is
\begin{ceqn}
\begin{equation}
    \vec{r}_{n}=\underbrace{x_{n}^{(0)} \hat{e}_{x}+y_{n}^{(0)} \hat{e}_{y}+z_{n}^{(0)} \hat{e}_{z}}_{\vec{r}_{n}^{(0)}}+\underbrace{x_{s} \sin \left(\frac{2 \pi x_{n}^{(0)}}{\lambda}\right) \hat{e}_{x}}_{\Delta \vec{r}_{n}},
\end{equation}
\end{ceqn}
where the amplitude of motion of the volume element follows the sinusoidal mode profile. We assume that the material is homogeneous, allowing us to approximate the density of the material as a constant $\rho\left(\vec{r}_{n}\right) \approx \rho_{0}$ and integrate over $\vec{r}^{(0)}$ instead of $\vec{r}$. Therefore, the expression for the mass density operator becomes
\begin{ceqn}
\begin{equation}
\begin{array}{r}m(\vec{s})=\frac{\rho_{0} r_{c}^{3}}{(2 \pi)^{3 / 2}} \int \mathrm{d}^{3} k e^{-\frac{1}{2} r_{c}^{2} k^{2}} e^{i \vec{k} \cdot \vec{s}} \cdot \int_{-\lambda / 4}^{\lambda / 4} e^{-i k_{x}\left(x^{(0)}+x_{s} \sin \left(\frac{2 \pi x_{n}^{(0)}}{\lambda}\right)\right)} \mathrm{d} x^{(0)} \\ \cdot \int_{-R}^{R} \int_{-\sqrt{R^{2}-y^{2}}}^{\sqrt{R^{2}-y^{2}}} \exp \left(-i k_{y} y_{n}^{(0)}\right) \exp \left(-i k_{z} z_{n}^{(0)}\right) \mathrm{d} z_{n}^{(0)} \mathrm{d} y_{n}^{(0)}.\end{array}
\end{equation}
\end{ceqn}
Here we have broken up the 3-D integral into a 1-D integral in the longitudinal coordinate $x_n^{(0)}$, and a 2-D integral in the transverse coordinates $y_n^{(0)}$ and $z_n^{(0)}$. The 2-D integral in transverse coordinates can be more easily solved by converting to polar coordinates, which we evaluate to be
\begin{ceqn}
\begin{equation}
\begin{split}
    \tilde{\rho}_{\perp}\left(k_{\perp}\right) = \tilde{\rho}_{\perp}\left(k_{y}, k_{z}\right) &= \int_{-R}^{ R}\int_{-\sqrt{R^2 - y^2}}^{\sqrt{R^2 - y^2}} \exp(-i k_y y_n^{(0)}) \exp(-i k_z z_n^{(0)}) \mathrm{d} z^{(0)}_n \mathrm{d} y^{(0)}_n \\
    &= \int_0^{2 \pi} \int_0^{R} r \exp(-i k_y r\sin(\theta)) \exp(-i k_z r\cos(\theta)) \mathrm{dr}\mathrm{d\theta} \\
    &= \frac{2 \pi R}{k_{\perp}}\mathrm{J}_1(R k_{\perp} ).
\end{split}
\end{equation}
\end{ceqn}
Here, $\mathrm{J}_{\alpha}$ are Bessel functions of the first kind. After making the substitution $u = \frac{2 \pi}{\lambda}x_n^{(0)}$, the longitudinal integral simplifies to
\begin{ceqn}
\begin{equation}
    \tilde{\rho}_{x}\left(k_{x}\right) = \left(\frac{\lambda}{2 \pi} \right) \cdot \int_{-\pi/ 2}^{\pi / 2} e^{-i k_{x} \left(\frac{\lambda u}{2 \pi}+ x_{s} \sin(u)\right)} \mathrm{d} u.
\end{equation}
\end{ceqn}
we approximate $\sin(x)$ as a quadratic for $0 \leq x \leq \pi$:
\begin{ceqn}
\begin{equation}
   \sin(x) \approx -\frac{4}{\pi^2}x^2 + \frac{4}{\pi}x,
\end{equation}
\end{ceqn}
while for $-\pi \leq x \leq 0$:
\begin{ceqn}
\begin{equation}
   \sin(x) \approx \frac{4}{\pi^2}x^2 + \frac{4}{\pi}x.
\end{equation}
\end{ceqn}
After making these approximations, the longitudinal integral becomes
\begin{ceqn}
\begin{equation}
    \tilde{\rho}_{x}\left(k_{x}\right) = \left(\frac{\lambda}{2 \pi} \right) \cdot \left( \int_{0}^{\pi / 2} e^{-i k_{x} \left(\frac{\lambda u}{2 \pi}+ x_{s} (-\frac{4}{\pi^2}u^2 + \frac{4}{\pi}u)\right)} \mathrm{d} u + \int_{-\pi/2}^{0} e^{-i k_{x} \left(\frac{\lambda u}{2 \pi}+ x_{s} (\frac{4}{\pi^2}u^2 + \frac{4}{\pi}u)\right)} \mathrm{d} u\right).
\end{equation}
\end{ceqn}
Now expressing the complex exponential $e^{-i k_{x} \left( x_{s} (-\frac{4}{\pi}u^2 + \frac{4}{\pi}u)\right)}$ in terms of $\mathrm{sin}$ and $\mathrm{cos}$, taking the Taylor series of the trigonometric funtions and eliminating all terms up to first order in $k_x x_s$ we obtain:
\begin{ceqn}
\begin{equation}
     \exp\left(-i k_{x} \left( x_{s} (-\frac{4}{\pi}u^2 + \frac{4}{\pi}u)\right)\right) \approx 1 - i k_x x_s\left( -\frac{4}{\pi^2}u^2 + \frac{4}{\pi}u\right).
\end{equation}
\end{ceqn}
After a similar calculation for $e^{-i k_{x} \left( x_{s} (\frac{4}{\pi^2}u^2 + \frac{4}{\pi}u)\right)}$ we obtain:
\begin{ceqn}
\begin{equation}
    \exp\left(-i k_{x} \left( x_{s} (\frac{4}{\pi}u^2 + \frac{4}{\pi}u)\right)\right) \approx 1 - i k_x x_s\left( \frac{4}{\pi^2}u^2 + \frac{4}{\pi}u\right).
\end{equation}
\end{ceqn}
We now substitute these expressions into the mass density operator:
\begin{ceqn}
\begin{equation}
\begin{split}
     m(\vec{s})=\frac{\rho_{0} r_{c}^{3}}{(2 \pi)^{3 / 2}} &\int \mathrm{d}^{3} k e^{-\frac{1}{2} r_{c}^{2} k^{2}} e^{i \vec{k} \cdot \vec{s}} \left(\frac{\lambda}{2 \pi} \right) \\ &\cdot ( \int_{0}^{\pi / 2} e^{-i k_{x} \left(\frac{\lambda u}{2 \pi}\right)} \left( 1 - i k_x x_s\left( -\frac{4}{\pi^2}u^2 + \frac{4}{\pi}u\right) \right) \mathrm{d} u \\ &+ \int_{-\pi/2}^{0} e^{-i k_{x} \left(\frac{\lambda u}{2 \pi}\right)} \left( 1 - i k_x x_s\left( \frac{4}{\pi^2}u^2 + \frac{4}{\pi}u\right) \right) \mathrm{d} u)\tilde{\rho}_{\perp}\left(k_{y}, k_z  \right) .
\end{split}
\end{equation} 
\end{ceqn}
Once substituted into the CSL master equation, only terms proportional to $x_s$ are non-zero, allowing the mass density operator to be expressed as
\begin{ceqn}
\begin{equation}
\begin{split}
     m(\vec{s})=\frac{\rho_{0} r_{c}^{3}}{(2 \pi)^{3 / 2}} &\int \mathrm{d}^{3} k e^{-\frac{1}{2} r_{c}^{2} k^{2}} e^{i \vec{k} \cdot \vec{s}} \left(\frac{2\lambda}{ \pi^2} \right) \\ &\cdot ( \int_{0}^{\pi / 2} e^{-i k_{x} \left(\frac{\lambda u}{2 \pi}\right)} \left( - i k_x x_s\left( -\frac{1}{\pi}u^2 + u\right) \right) \mathrm{d} u \\&+ \int_{-\pi/2}^{0} e^{-i k_{x} \left(\frac{\lambda u}{2 \pi}\right)} \left(  - i k_x x_s\left( \frac{1}{\pi}u^2 + u\right) \right) \mathrm{d} u)\tilde{\rho}_{\perp}\left(k_{y}, k_z  \right) .
\end{split}
\end{equation} 
\end{ceqn}
We can now express the mass density operator as
\begin{ceqn}
\begin{equation}
    m(\vec{s})  =\frac{r_{\mathrm{csl}}^{3}\rho_0}{(2 \pi)^{3 / 2}} \int \mathrm{d}^{3} \vec{k} e^{-\frac{1}{2}r_{\mathrm{csl}}^{2} k^{2}} e^{i \vec{k} \cdot \vec{s}} \tilde{\rho}_{x}\left(k_{x}\right) \tilde{\rho}_{\perp}\left(k_{\perp}\right),
\end{equation}
\end{ceqn}
substituting this expression into the CSL master equation, it becomes:
\begin{ceqn}
\begin{equation}
  \mathcal{L}_{\mathrm{csl}} \hat{\rho} \approx  \frac{\lambda_{\mathrm{csl}} r_{\mathrm{csl}}^{3} \rho_{0}^{2}}{\pi^{3 / 2} \mathrm{amu}^{2}} \int \mathrm{d}^{3} k e^{-r_{\mathrm{csl}}^{2} k^{2}} \left|\tilde{\rho}_{\perp}\right|^{2} \left(\tilde{\rho}_{x} \hat{\rho} \tilde{\rho}_{x}-\frac{1}{2}\left\{\hat{\rho},\left|\tilde{\rho}_{x}\right|^{2}\right\}\right),
\end{equation}
\end{ceqn}
here we have applied $\int e^{i\left(\vec{k}-\vec{k}^{\prime}\right) \cdot \vec{s}} \mathrm{d}^3 \vec{s} = (2 \pi)^{3} \delta\left(\vec{k}-\vec{k}^{\prime}\right)$ to reduce the number of variables. Simplifying with the identity
\begin{ceqn}
\begin{equation}
    \tilde{\rho}_{x} \hat{\rho} \tilde{\rho}_{x}-\frac{1}{2}\left\{\hat{\rho},\left|\tilde{\rho}_{x}\right|^{2}\right\}=-\frac{1}{2}\left[\tilde{\rho}_{x},\left[\tilde{\rho}_{x}, \hat{\rho}\right]\right],
\end{equation}
\end{ceqn}
and expressing the commutator in terms of the position operator for the displacement
\begin{ceqn}
\begin{equation}
    -\frac{1}{2}\left[\tilde{\rho}_{x},\left[\tilde{\rho}_{x}, \hat{\rho}\right]\right] =  |\tilde{\rho}'_{x}\left(k_{x}\right)|^2 \cdot (-1) \cdot \left[x_{s},\left[x_{s}, \hat{\rho}\right]\right]
\end{equation}
\end{ceqn}
where
\begin{ceqn}
\begin{equation}
    |\tilde{\rho}'_{x}\left(k_{x}\right)|^2 =\left(\frac{\lambda}{2 \pi} \right)^2 k_x^2 \frac{\left(-8+\left(8+a^{2} \pi^{2}\right) \cos \left(\frac{a \pi}{2}\right)\right)^{2}}{4 a^{6} \pi^{2}},
\end{equation}
\end{ceqn}
and $a = k_x \lambda/2\pi $, we can finally express the CSL Master equation as
\begin{ceqn}
\begin{equation}
\begin{split}
    \mathcal{L}_{\mathrm{csl}} \hat{\rho}&= -\frac{\lambda_{c} r_{c}^{3} \rho_{0}^{2}}{\pi^{3 / 2} \mathrm{amu}^{2}} \left(\frac{16}{\pi \lambda}\right) \int \exp(-r_c^2 \left(\frac{2\pi a}{\lambda}\right)^2)\frac{\left(-8+\left(8+a^{2} \pi^{2}\right) \cos \left(\frac{a \pi}{2}\right)\right)^{2}}{4 a^{4} \pi^{2}} \mathrm{d}a\\  &\cdot \int \int e^{-r_{\mathrm{csl}}^{2} (k_{x}^{2} + k_y^2)} \frac{4 \pi^2 R^2}{(k_y^2 + k_z^2)} \mathrm{J}_{1}^2\left(R \sqrt{k_y^2 + k_z^2}\right)\mathrm{d}k_y \mathrm{d}k_z (-1)\left[x_{s},\left[x_{s}, \hat{\rho}\right]\right].
\end{split}
\end{equation}
\end{ceqn}
We evaluate the 2-D integral in transverse coordinates to be \cite{Nimmrichter2014}:
\begin{ceqn}
\begin{equation}
\begin{split}
    &\iint e^{-r_{\mathrm{csl}}^{2}\left(k_{\mathrm{x}}^{2}+k_{y}^{2}\right)} \frac{4 \pi^{2} R^{2}}{\left(k_{y}^{2}+k_{z}^{2}\right)} \mathrm{J}_{1}^{2}\left(R \sqrt{k_{y}^{2}+k_{z}^{2}}\right) \mathrm{d} k_{y} \mathrm{d} k_{z} \\
    &= 4\pi^3 R^2 \left[\frac{\exp(\frac{R^2}{2 r_{\mathrm{CSL}}^2}) - \mathrm{I}_0\left(\frac{R^2}{2 r_{\mathrm{CSL}}^2} \right) - \mathrm{I}_1\left(\frac{R^2}{2 r_{\mathrm{CSL}}^2} \right) }{2\exp(\frac{R^2}{2 r_{\mathrm{CSL}}^2})} \right].
    \end{split}
\end{equation}
\end{ceqn}
Here $\mathrm{I}_n$ are the modified Bessel functions of the first kind. Therefore, the CLS-Lindblad operator is

\begin{ceqn}
\begin{equation}
    \begin{split}
         \mathcal{L}_{\mathrm{csl}} \hat{\rho}= - \lambda_c \eta_{\mathrm{new}} \left[x_{s},\left[x_{s}, \hat{\rho}\right]\right].
    \end{split}
\end{equation}
\end{ceqn}
where
\begin{ceqn}
\begin{equation} \label{etanew}
\begin{split}
  \eta_{\mathrm{new}} =  \frac{ r_{c}^{3} \rho_{0}^{2}R^2}{ \mathrm{amu}^{2}} \left(\frac{64 \pi^{1/2}}{ \lambda}\right) &\cdot \left[\frac{\exp \left(\frac{R^{2}}{2 r_{\mathrm{c}}^{2}}\right)-\mathrm{I}_{0}\left(\frac{R^{2}}{2 r_{\mathrm{c}}^{2}}\right)-\mathrm{I}_{1}\left(\frac{R^{2}}{2 r_{\mathrm{c}}^{2}}\right)}{2 \exp \left(\frac{R^{2}}{2 r_{\mathrm{c}}^{2}}\right)}\right] \\
         &\cdot \int_{-\infty}^{\infty} \exp(-r_c^2 \left(\frac{2\pi a}{\lambda}\right)^2)\frac{\left(-8+\left(8+a^{2} \pi^{2}\right) \cos \left(\frac{a \pi}{2}\right)\right)^{2}}{4 a^{4} \pi^{2}} \mathrm{d}a,
\end{split}
\end{equation}
\end{ceqn}

We now take the mode expansion for $x_s$ by substituting $x_s = x_0\left(b + b^{\dag}\right)$ where $x_0 = \sqrt{\frac{\hbar}{2 m_{\mathrm{eff}} \omega}}$, such that $\tilde{D} = \lambda_c x_0^2  \eta_{\mathrm{new}}$ is the CSL-induced decoherence rate. We can neglect terms of the form $\hat{b}^{\dag}\hat{b}^{\dag}$ and $\hat{b}\hat{b}$ as long as the mechanical oscillator's resonance frequency $\omega$ is much larger than the decoherence rate $\tilde{D}$ and the oscillator's damping rate $\gamma_r$: $\tilde{D}, \gamma_r \ll \omega$. In this regime after taking the mode expansion, the CSL-Lindblad term has the form $-\lambda_c \eta_{\mathrm{new}}\left[\delta_{x},\left[\delta_{x}, \hat{\rho}\right]\right] \approx 2 \tilde{D} (\mathcal{D}[\hat{b}] + \mathcal{D}[\hat{b}^{\dag}] )\hat{\rho}$. If the mechanical oscillator is in its ground state only the $\mathcal{D}[\hat{b}^{\dag}]$ term applies. The coefficient of the $\mathcal{D}[\hat{b}^{\dag}]$ superoperator quantifies the rate at which phonons are generated in the resonator.  

We now define the dimensionless the dimensionless CSL cross-section of a cylindrical bulk acoustic wave resonator to be:
\begin{ceqn}
\begin{equation}
\begin{split}
    D &= \frac{2\tilde{D}}{\lambda_c}, \\
    &= 2 x_0^2 \eta_{\mathrm{new}}.
\end{split}
\end{equation}
\end{ceqn}
For Silicon Carbide (SiC) with a density of $\rho_0 \approx 3210 \; \mathrm{kg m^{-3}}$, and a cylindrical resonator of radius $R \approx 35 \mathrm{\mu m}$ with a phonon wavelength of $\lambda \approx 15 \times 10^{-7} \; \mathrm{\ m}$, which corresponds to a mode frequency of $\omega/2 \pi \approx 6.33\; \mathrm{GHz}$ and assuming the expected collapse noise correlation length to be $r_c \approx 10^{-7} \; \mathrm{ m}$, we obtain
\begin{ceqn}
\begin{equation} \label{crossaseccca}
    D \approx 7.8 \times 10^{5}
\end{equation}
\end{ceqn}
at $r_c = 10^{-7} \; \mathrm{m}$.

Eq.~\eqref{etanew} shows that the CSL cross-section is maximised for phonon wavelengths $\lambda \approx 6 r_c$. Therefore, a phonon wavelength of $\lambda \approx 15 \times 10^{-7} \; \mathrm{\ m}$ is chosen to enhance the sensitivity to collapse-induced heating, while ensuring that the mode frequency is within the tunable frequency range of the qubit.

\subsubsection{Generalising to multiple anti-nodes}
Here we generalise the calculation of the CSL cross-section to multiple anti-nodes. The Gaussian-smeared mass density operator for CSL is defined as
\begin{ceqn}
\begin{equation}
    m(\vec{s})=\sum_{n} m_{n} e^{-\frac{(\vec{s}-\vec{r} n)^{2}}{2 r_{c}^2}}=\int_{V} \mathrm{~d}^{3} \vec{r} \rho(\vec{r}) e^{-\frac{(\vec{s}-\vec{r})^{2}}{2 r_{c}^{2}}},
\end{equation}
\end{ceqn}
where $\vec{r}_n$ is the coordinate of the $n$-th mass $m_n$ and $\rho(\vec{r})$ is the mass density. By taking a Fourier transform, the mass density can be expressed as
\begin{ceqn}
\begin{equation} \label{massdensch6}
    m(\vec{s})=\frac{r_{c}^{3}}{(2 \pi)^{3 / 2}} \int \mathrm{d}^{3} \vec{k} e^{-\frac{1}{2} r_{c}^{2} k^{2}} \int_{V} \mathrm{~d}^{3} \vec{r}_{n} \rho\left(\vec{r}_{n}\right) e^{i \vec{k} \cdot \left(\vec{s}-\vec{r}_{n}\right)}.
\end{equation}
\end{ceqn}
The position operator for an infinitesimal volume element of the continuous structure is 
\begin{ceqn}
\begin{equation}
    \vec{r} = \vec{r}_{\mathrm{CoM}} + \vec{r}^{(0)} + \Delta \vec{r}, 
\end{equation}
\end{ceqn}
where $\Delta \vec{r}$ is the displacement of the volume element relative to its equilibrium position, $\vec{r}^{(0)}$ is the equilibrium position of the volume element relative to the cenre of mass and $\vec{r}_{\mathrm{CoM}}$ is the centre-of-mass (CoM) operator. For an internal mechanical mode of a resonator there is no oscillation of the system's centre-of-mass requiring meaning that $\vec{r}_{\mathrm{CoM}} = 0$. Therefore, we must examine more closely the effects of $\Delta \vec{r}$. In this work, we will model $\Delta \vec{r}$ as an ideal \textit{breathing mode}.

Now, we approximate the density as constant ($\rho = \rho_0$), and integrate over the length of a resonator with $l +1$ antinodes, which has length $h = (l+1) \frac{\lambda}{2}$. In this case, the expression for the mass density operator is
\begin{equation} \label{massdensch6}
    m(\vec{s})=\frac{r_{c}^{3}}{(2 \pi)^{3 / 2}} \int \mathrm{d}^{3} \vec{k} e^{-\frac{1}{2} r_{c}^{2} k^{2}} \int_{- \frac{\lambda}{4}}^{(l+ \frac{1}{2}) \frac{\lambda}{2}} \mathrm{~d}^{3} \vec{r}_{n} \rho_0 e^{i \vec{k} \cdot \left(\vec{s}-\vec{r}_{n}\right)},
\end{equation}
where the position operator of a volume element in the 1-D breathing mode is
\begin{equation}
    \vec{r}_{n}=\underbrace{x_{n}^{(0)} \hat{e}_{x}+y_{n}^{(0)} \hat{e}_{y}+z_{n}^{(0)} \hat{e}_{z}}_{\vec{r}_{n}^{(0)}}+\underbrace{x_{s} \sin \left(\frac{2 \pi x_{n}^{(0)}}{\lambda}\right) \hat{e}_{x}}_{\Delta \vec{r}_{n}}.
\end{equation}
Now, the sum in the mass density operator can be broken up into l parts:
\begin{equation}
    \begin{aligned}
m_{\text {new }}(\vec{s}) &=\sum_{j=0}^{j=l} \frac{r_{\mathrm{csl}}^{3}}{(2 \pi)^{3 / 2}} \int \mathrm{d}^{3} \vec{k} e^{i \vec{k} \cdot \vec{s}} e^{-\frac{1}{2} r_{\mathrm{csl}}^{2} k^{2}} \int_{\left(j-\frac{1}{2}\right) \frac{\lambda}{2}}^{\left(j+\frac{1}{2}\right) \frac{\lambda}{2}} \exp \left(-i k_{x}\left(x_{n}^{(0)}+x_{s} \sin \left(\frac{2 \pi}{\lambda}x_n^{(0)}\right)\right)\right) \mathrm{d} x_{n}^{(0)} \\
& \cdot \int_{-R}^{R} \int_{-\sqrt{R^{2}-y^{2}}}^{\sqrt{R^{2}-y^{2}}} \exp \left(-i k_{y} y_{n}^{(0)}\right) \exp \left(-i k_{z} z_{n}^{(0)}\right) \mathrm{d} z_{n}^{(0)} \mathrm{d} y_{n}^{(0)}.
\end{aligned}
\end{equation}
Now, we have:
\begin{equation}
\begin{split}
    \tilde{\rho}_{\perp}\left(k_{\perp}\right)=\tilde{\rho}_{\perp}\left(k_{y}, k_{z}\right)&=\int_{-R}^{R} \int_{-\sqrt{R^{2}-y^{2}}}^{\sqrt{R^{2}-y^{2}}} \exp \left(-i k_{y} y_{n}^{(0)}\right) \exp \left(-i k_{z} z_{n}^{(0)}\right) \mathrm{d} z_{n}^{(0)} \mathrm{d} y_{n}^{(0)} \\
    &= \frac{2 \pi R}{k_{\perp}} \mathrm{J}_{1}\left(R k_{\perp}\right).
\end{split}
\end{equation}
Now, after making the substitution $x_n^{(0)} \rightarrow x_n^{(0)} + \frac{j \lambda}{2}$ and then the substitution $u=\frac{2 \pi}{\lambda} x_{n}^{(0)}$, the longitudinal integral simplifies to
\begin{equation}
    \tilde{\rho}_{x}\left(k_{x}\right)= \sum_{j=0}^{j=l} \left(\frac{\lambda}{2 \pi}\right) \cdot \int_{-\pi / 2}^{\pi / 2} e^{-i k_{x}\left(\frac{\lambda u}{2 \pi} +\left( \frac{j\lambda}{2} \right) + (-1)^{j} x_{s} \sin (u)\right)} \mathrm{d}u.
\end{equation}
After approximating the sin term as a quadratic, and taking the expansion for small $x_s$, we have:
\begin{equation}
\begin{aligned}
   \tilde{\rho}_{x}\left(k_{x}\right) =  \sum_{j=0}^{j=l}  & \int \mathrm{d}^{3} k e^{-\frac{1}{2} r_{c}^{2} k^{2}} e^{i \vec{k} \cdot \vec{s}}\left(\frac{\lambda}{2 \pi}\right) e^{-i k_x j \frac{\lambda}{2}} \\
& \cdot\left(\int_{0}^{\pi / 2} e^{-i k_{x}\left(\frac{\lambda u}{2 \pi}\right)}\left(1 + (-1)^{j} i k_{x} x_{s}\left(-\frac{4}{\pi^{2}} u^{2}+\frac{4}{\pi} u\right)\right) \mathrm{d} u\right.\\
&\left.+\int_{-\pi / 2}^{0} e^{-i k_{x}\left(\frac{\lambda u}{2 \pi}\right)}\left(1 + (-1)^{j} i k_{x} x_{s}\left(\frac{4}{\pi^{2}} u^{2}+\frac{4}{\pi} u\right)\right) \mathrm{d} u\right).
\end{aligned}
\end{equation}
Substituting this into the mass density operator, we obtain:

\begin{equation}
   \begin{aligned}
m(\vec{s})=\frac{\rho_{0} r_{c}^{3}}{(2 \pi)^{3 / 2}} \sum_{j=0}^{j=l}  & \int \mathrm{d}^{3} k e^{-\frac{1}{2} r_{c}^{2} k^{2}} e^{i \vec{k} \cdot \vec{s}}\left(\frac{\lambda}{2 \pi}\right) e^{-i k_x j \frac{\lambda}{2}} \\
& \cdot\left(\int_{0}^{\pi / 2} e^{-i k_{x}\left(\frac{\lambda u}{2 \pi}\right)}\left(1 + (-1)^j i k_{x} x_{s}\left(-\frac{4}{\pi^{2}} u^{2}+\frac{4}{\pi} u\right)\right) \mathrm{d} u\right.\\
&\left.+\int_{-\pi / 2}^{0} e^{-i k_{x}\left(\frac{\lambda u}{2 \pi}\right)}\left(1 + (-1)^j i k_{x} x_{s}\left(\frac{4}{\pi^{2}} u^{2}+\frac{4}{\pi} u\right)\right) \mathrm{d} u\right) \tilde{\rho}_{\perp}\left(k_{y}, k_{z}\right) .
\end{aligned}
\end{equation}
We can now effectively express the mass density operator as
\begin{ceqn}
\begin{equation}
    m(\vec{s})  = \sum_{j = 0}^{j = l}\frac{r_{\mathrm{csl}}^{3}\rho_0}{(2 \pi)^{3 / 2}} \int \mathrm{d}^{3} \vec{k} e^{-\frac{1}{2}r_{\mathrm{csl}}^{2} k^{2}} e^{i \vec{k} \cdot \vec{s}} e^{- i k_x j \frac{\lambda}{2}}  \tilde{\rho}'_{x}\left(k_{x}\right) \tilde{\rho}_{\perp}\left(k_{\perp}\right) (-1)^j,
\end{equation}
\end{ceqn}
where $\tilde{\rho}'_{x}$ is the term in $\tilde{\rho}_{x}$ proportional to $x_s$ with terms dependent on $j$ factored out. Substituting this expression into the CSL master equation, it becomes:
\begin{ceqn}
\begin{equation}
  \mathcal{L}_{\mathrm{csl}} \hat{\rho} \approx  \frac{\lambda_{\mathrm{csl}} r_{\mathrm{csl}}^{3} \rho_{0}^{2}}{\pi^{3 / 2} \mathrm{amu}^{2}} \int \mathrm{d}^{3} k e^{-r_{\mathrm{csl}}^{2} k^{2}} \left|\tilde{\rho}_{\perp}\right|^{2} \left(\tilde{\rho}_{x} \hat{\rho} \tilde{\rho}_{x}-\frac{1}{2}\left\{\hat{\rho},\left|\tilde{\rho}_{x}\right|^{2}\right\}\right),
\end{equation}
\end{ceqn}
here we have applied $\int e^{i\left(\vec{k}-\vec{k}^{\prime}\right) \cdot \vec{s}} \mathrm{d}^3 \vec{s} = (2 \pi)^{3} \delta\left(\vec{k}-\vec{k}^{\prime}\right)$ to reduce the number of variables. Simplifying with the identity
\begin{ceqn}
\begin{equation}
    \tilde{\rho}_{x} \hat{\rho} \tilde{\rho}_{x}-\frac{1}{2}\left\{\hat{\rho},\left|\tilde{\rho}_{x}\right|^{2}\right\}=-\frac{1}{2}\left[\tilde{\rho}_{x},\left[\tilde{\rho}_{x}, \hat{\rho}\right]\right],
\end{equation}
\end{ceqn}
and expressing the commutator in terms of the position operator for the displacement
\begin{ceqn}
\begin{equation}
    -\frac{1}{2}\left[\tilde{\rho}_{x},\left[\tilde{\rho}_{x}, \hat{\rho}\right]\right] =  |\tilde{\rho}'_{x}\left(k_{x}\right)|^2 \cdot (-1) \cdot \left[x_{s},\left[x_{s}, \hat{\rho}\right]\right]
\end{equation}
\end{ceqn}
where
\begin{ceqn}
\begin{equation} \label{rhoxdash}
    |\tilde{\rho}'_{x}\left(k_{x}\right)|^2 = \sum_{j = 0}^{j = l} \sum_{j' = 0}^{j' = l}\left(\frac{\lambda}{2 \pi} \right)^2 (-1)^{(j+ j')} e^{-i k_x(j - j')\frac{\lambda}{2}}k_x^2 \frac{\left(-8+\left(8+a^{2} \pi^{2}\right) \cos \left(\frac{a \pi}{2}\right)\right)^{2}}{4 a^{6} \pi^{2}},
\end{equation}
\end{ceqn}
and $a = k_x \lambda/2\pi $. Now, we can further simplify this expression, by evaluating the double sum over $j$ and $j'$. Here we also note that for a single node, such that there is only one term in the sum, the factor $e^{-i k_x(j + j')\frac{\lambda}{2}}$ disappears, and we obtain the expression from our calculation for a single node.
The sum
\begin{ceqn}
\begin{equation}
    \sum_{j = 0}^{j = l} (-1)^j e^{-i k_x(j)\frac{\lambda}{2}}
\end{equation}
\end{ceqn}
is a geometric series which evaluates to
\begin{ceqn}
\begin{equation}
    \frac{(1 - e^{-ik_x(l + 1)\frac{\lambda}{2}})}{(1 + e^{-ik_x\frac{\lambda}{2}})},
\end{equation}
\end{ceqn}
assuming $l + 1$ is even. 

After some algebra, we find that the integral in the expression from the previous section is modified to include frequency phases to the sinusoidal terms.  

For $l + 1 = 40$, eq.~\eqref{rhoxdash} simplifies to
\begin{ceqn}
\begin{equation}
    \left|\tilde{\rho}_{x}^{\prime}\left(k_{x}\right)\right|^{2} = \left(\frac{\lambda}{2 \pi}\right)^{2} k_{x}^{2} \frac{\left(-8+\left(8+a^{2} \pi^{2}\right) \cos \left(\frac{a \pi}{2}\right)\right)^{2}}{4 a^{6} \pi^{2}} \frac{4 e^{i \pi a} \sin^2(20a\pi)}{(1 + e^{i a \pi})^2},.
\end{equation}
\end{ceqn}
The CSL Linblad operator is
\begin{ceqn}
\begin{equation}
    \mathcal{L}_{\mathrm{csl}} \hat{\rho}=-\lambda_{c} \eta_{\text {new }}\left[x_{s},\left[x_{s}, \hat{\rho}\right]\right] ,
\end{equation}
\end{ceqn}
where,
\begin{equation}
    \begin{aligned}
\eta_{\text {new }}=&\frac{r_{c}^{3} \rho_{0}^{2} R^{2}}{a m u^{2}}\left(\frac{64 \pi^{1 / 2}}{\lambda}\right)  \cdot\left[\frac{\exp \left(\frac{R^{2}}{2 r_{c}^{2}}\right)-\mathrm{I}_{0}\left(\frac{R^{2}}{2 r_{c}^{2}}\right)-\mathrm{I}_{1}\left(\frac{R^{2}}{2 r_{c}^{2}}\right)}{2 \exp \left(\frac{R^{2}}{2 r_{c}^{2}}\right)}\right] \\
& \cdot \int_{-\infty}^{\infty} \exp \left(-r_{c}^{2}\left(\frac{2 \pi a}{\lambda}\right)^{2}\right) \frac{\left(-8+\left(8+a^{2} \pi^{2}\right) \cos \left(\frac{a \pi}{2}\right)\right)^{2}}{4 a^{4} \pi^{2}} \frac{4 e^{i \pi a} \sin^2(20 \pi a)}{(1 + e^{i \pi a})^2}\mathrm{~d} a
\end{aligned}
\end{equation}
Expressing in terms of the relevant integration variables, we obtain:
\begin{equation}
\begin{aligned}
    \eta_{\text {new }}=&\frac{r_{c}^{3} \rho_{0}^{2} R^{2}}{a m u^{2}}\left(\frac{64 \pi^{1 / 2}}{\lambda}\right)  \cdot\left[\frac{\exp \left(\frac{R^{2}}{2 r_{c}^{2}}\right)-\mathrm{I}_{0}\left(\frac{R^{2}}{2 r_{c}^{2}}\right)-\mathrm{I}_{1}\left(\frac{R^{2}}{2 r_{c}^{2}}\right)}{2 \exp \left(\frac{R^{2}}{2 r_{c}^{2}}\right)}\right] \\
& \cdot \int_{-\infty}^{\infty} \exp \left(-r_{c}^{2}\left(\frac{2 \pi a}{\lambda}\right)^{2}\right) \frac{\left(-8+\left(8+a^{2} \pi^{2}\right) \cos \left(\frac{a \pi}{2}\right)\right)^{2}}{4 a^{4} \pi^{2}} \left( \frac{4 e^{i \pi a} \sin^2(20 \pi a)}{(1 + e^{i \pi a})^2} \right) \mathrm{~d} a.
\end{aligned}
\end{equation}
Numerically evaluating the integral, at the expected correlation length, we obtain the following result for $l + 1 = 40$:
\begin{ceqn}
\begin{equation}
    D = x_0^2 \eta_{\mathrm{new}} \approx 5.5 \times 10^5.
\end{equation}
\end{ceqn}

\section{Measurement Scheme}
\begin{figure}[h] 
\begin{center}
\includegraphics[width = 160mm, height = 80mm]{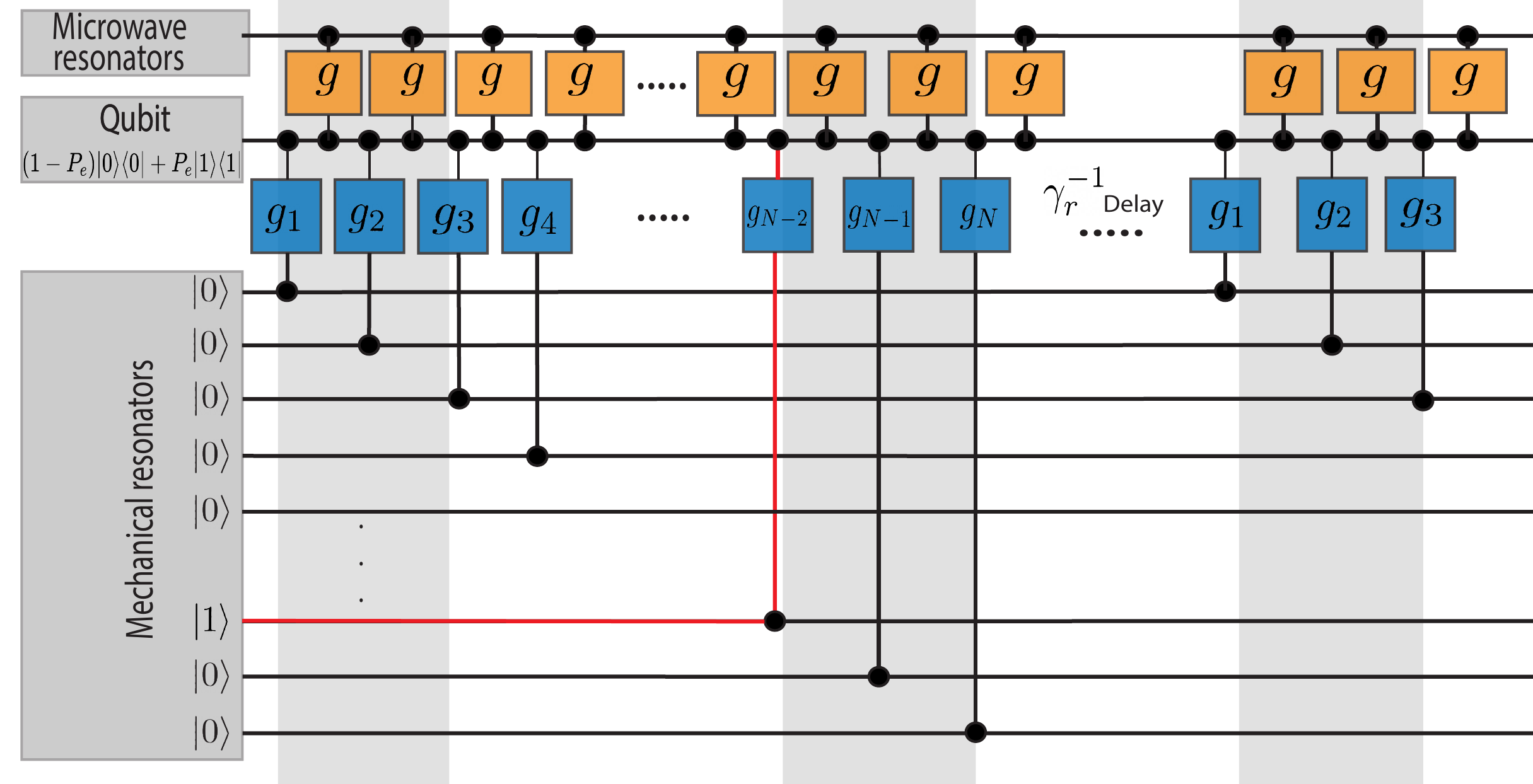}
    \caption{\label{expimpfig} Experimental protocol. Swap operations are sequentially performed between the qubit and an array of mechanical resonances, with mechanical resonator $j$ having an electomechanical coupling rate $g_j$. The qubit state is measured after each swap via interaction with a microwave cavity with coupling rate $g$. The first few measurements, both at the beginning of the protocol and after a detection event (red lines) are discarded, with the associated swap operations instead used to initialize the qubit to near its ground state. After all mechanical resonators have been measured, the qubit is decoupled for a time of $\sim\gamma_r^{-1} \; \mathrm{s}$ in order to allow the resonators to cool and collapse-induced heating to accumulate. The protocol is then repeated until sufficient measurement statistics have been obtained.}
\end{center}
\end{figure}

As shown in Figure~\ref{expimpfig}, after the first swap operation and dispersive measurement, the qubit is then tuned on-resonance with the second mechanical resonator on a timescale fast compared to the swap operation period. This is followed by another swap operation and dispersive measurement. This process is repeated for all mechanical modes in the array. Once all mechanical modes  have had one interaction with the qubit, the qubit is then decoupled from all resonators for $T \sim \gamma_r^{-1} \; \mathrm{s}$ in order to allow the resonators to cool and collapse-induced heating to re-accumulate. This process is repeated to build up statistics. A given CSL collapse rate can be ruled out if it predicts a higher average phonon occupation than is observed for the relevant correlation length.

\section{Interaction Hamiltonian and coupling strength} \label{coupleappendix}

The interaction energy between the qubit and the BAW resonator is given by
\begin{ceqn}
\begin{equation} \label{haminteraction}
    H=-\int \sigma(\vec{x}) s(\vec{x}) d\mathrm{V}.
\end{equation}
\end{ceqn}
Here, $\sigma(\vec{x})$ is the stress generated by the piezoelectric transducer, $E(\vec{x})$ is the qubit's electric field. Now, the stress generated by the AIN is,
\begin{ceqn}
\begin{equation}
    \sigma(\vec{x}) = c_{33}d_{33}(\vec{x})E(\vec{x}).
\end{equation}
\end{ceqn}
Here $d_{33}(\vec{x})$ is the dominant piezoelectric tensor component (component pointing in the longitudinal direction), $c_{33}$ is the dominant stiffness tensor component of the resonator. We approximate that only the dominant components of the stiffness and piezoelectric tensors contribute to the interaction energy.  Now, we quantise the qubit mode as $E(\vec{x}) (\hat{a} + \hat{a}^{\dag})$, and the phonon mode as $s(\vec{x}) (\hat{b} + \hat{b}^{\dag})$, we obtain:
\begin{ceqn}
\begin{equation}
    H = -c_{33}d_{33}\int E(\vec{x})(\hat{a} + \hat{a}^{\dag})s(\vec{x})(\hat{b} + \hat{b}^{\dag}) d V.
\end{equation}
\end{ceqn}
We have approximated $d_{33}(\vec{x})$ as a constant inside the transducer disk and zero everywhere else. 

Assuming longitudinally polarised thickness modes confined below the PZT transducer, the strain profile of the phonon mode is 

\begin{ceqn}
\begin{equation}
    s_{l, m}(\vec{x})=\beta_{l, m} \sin \left(\frac{l \pi z}{h}\right)s_{\perp}\left( r, \theta \right),
\end{equation}
\end{ceqn}
where $s_{\perp}\left( r, \theta \right)$ is the phonon's transverse mode profile and $\beta_{l,m}$ is a normalisation constant so that the total energy of the mode equal $\hbar\omega_{l,m}$. Here the mode frequency is given by
\begin{ceqn}
\begin{equation}
    \omega_{l, m}=\sqrt{\left(\frac{l \pi}{h}\right)^{2} v_{l}^{2}+\left(\frac{2 j_{0, m}}{d}\right)^{2} v_{t}^{2}},
\end{equation}
\end{ceqn}
and the normalisation factor is
\begin{ceqn}
\begin{equation}
    \beta_{l, m}=\sqrt{\frac{\hbar \omega_{l, m}}{\pi h c_{33} \int J_{0}^{2}\left(\frac{2 j_{0, m} r}{d}\right) r d r}}.
\end{equation}
\end{ceqn}

We now assume that the electric field is the constant $E(\vec{x}) = E_0$ throughout the transducer, and then apply the rotating wave approximation at the fundamental transverse mode $\omega_{l, 0}$ and equate the interaction energy to the Jaynes-Cummings Hamiltonian $H_{\mathrm{int}}~=~-\hbar g (\hat{a}\hat{b}^{\dag} + \hat{a}^{\dag}\hat{b} )$, where $g$ is the coupling strength between the qubit and the BAW resonator. This gives us the following expression for the coupling strength:
\begin{ceqn}
\begin{equation} \label{couple}
\begin{split}
    \hbar g &= c_{33}d_{33}E_0\beta_{l,m}\int \sin\left(\frac{l \pi}{h}z \right)s_{\perp}\left( r, \theta \right)d V \\
    &= c_{33}d_{33}E_0\beta_{l,m}\int_0^{h}\sin\left(\frac{l \pi}{h}z \right) \mathrm{d}z \\ &\cdot \int_0^{2 \pi}\int_0^{d_e/2} rs_{\perp}\left( r, \theta \right) \mathrm{d}r\mathrm{d}\theta 
\end{split}
\end{equation}
\end{ceqn}
Here  $d_e$ is the diameter of the qubit electrode which can be chosen to selectively couple to the fundamental transverse mode. The expression for the coupling strength in Eq.~\eqref{couple} is exactly true for a cylindrical flat-cap cavity geometry and it is an excellent approximation for a plano-convex cap geometry as long as the radius of curvature is large compared to the cavity height $R \gg h$, where $h$ height of the cylindrical cavity. 

Henceforth we will assume that the plano-convex radius of curvature is large such that we can approximate the transverse modes as Bessel functions. Using the Bessel function solutions for the cylindrical flat-cap cavity geometry \cite{Chu2017} $s_{\perp}\left( r, \theta \right) = J_{0}\left(\frac{2 j_{0, m} r}{d}\right)$, we obtain the following expression for the coupling strength:
\begin{ceqn}
\begin{equation}
    \hbar g_{l,m} = 2 \left(\frac{2h}{l} \right)c_{33}d_{33}E_0\beta_{l,m} \int_0^{d_e/2} J_{0}\left(\frac{2 j_{0, m} r}{d}\right) r \mathrm{dr}.
\end{equation}
\end{ceqn}

To take a specific example, targeting a correlation length on the order of $r_c \approx 10^{-7} \; \mathrm{m}$, we consider a BAW resonator internal mode with a frequency of $\omega/2\pi = 6.33 \; \mathrm{GHz}$ as have been previously coupled to a qubit in Ref.~\cite{Chu2017}. In Ref.~\cite{Chu2017}, the  $l = 503$ longitudinal and $m = 0$ transverse mode of a 420~$\mathrm{\mu m}$ thick sapphire BAW resonator was used. Here, we propose to use SiC instead of sapphire due to the ultra-high Q factors observed in SiC BAW resonators \cite{Gokhale2020}. We use parameters based on Ref.~\cite{Chu2017}, but a shorter resonator to allow a larger coupling strength. We propose to couple to the $l = 40$ longitudinal and $m = 0$ transverse mode,  which corresponds to a phonon wavelength of $\lambda \approx 15 \times 10^{-7}~\mathrm{m}$. As discussed earlier, in a BAW resonator the CSL cross-section is maximised for phonon wavelengths $\lambda \approx 6 r_c$, such that any resonator with $\lambda \sim r_c$ has a large CSL cross-section at $r_c$, however phonon wavelengths that are slightly larger than $r_c$ are more ideal, allowing our choice of phonon wavelength to target the correlation length $r_c \approx 10^{-7} \; \mathrm{m}$, while providing a resonance frequency within the tunable frequency range of the qubit. 

In contrast to recent experiments which used AIN as the transducer material \cite{Chu2017, Chu2018}, we propose to use PZT. Reasonably conservative estimates of the piezoelectricity of single crystal PZT (${d_{33} \approx 80 \; \mathrm{pm}/V}$) \cite{Frag2014} offer a significant improvement on the piezoelectricity of AIN (${d_{33} \approx 5 \; \mathrm{pm}/V}$), allowing us to estimate an improved coupling strength in our proposed experiment.

\begin{figure}[h] 
\begin{center}
\includegraphics[scale  = 0.7]{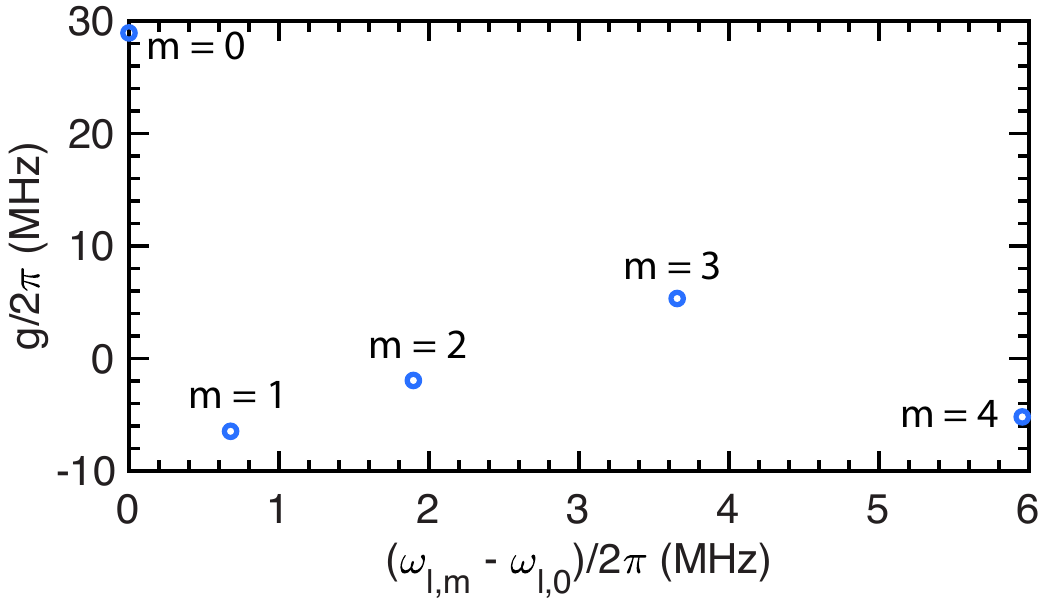}
    \caption{\label{couplestrength} Qubit-BAW resonator coupling strengths for the $l = 40$ longitudinal mode for a quartz resonator of height $h = 30 \; \mu \mathrm{m}$. The modes go to zero at the boundary of the cavity of diameter $d = 70 \; \mathrm{\mu m}$. The qubit electrode diameter $d_e = 55 \; \mathrm{\mu m}$ is slightly smaller to create selective coupling to the fundamental transverse mode.  }
\end{center}
\end{figure}

In Fig.~\ref{couplestrength} we show coupling strength as a function of mode frequency $\omega_{l,m}$ for a longitudinal mode number $l = 40$ and the lowest order transverse mode numbers of $m = 0$ to $m = 4$ for a cylindrical geometry bulk acoustic wave resonator. We use $d = 70 \; \mathrm{\mu m}$ for the cavity diameter, $d_e = 55 \mathrm{\mu m}$ for the electrode diameter, $E_0 = 2.8 \times 10^{-2} \; \mathrm{V/m}$, $c_{33} = 501 \; \mathrm{GPa}$ and a cavity vertical height of $h = 30 \; \mu\mathrm{m} $. With these parameters, we find the fundamental transverse mode at $l = 40$ has frequency $\omega_{l, 0}/2\pi = 6.33 \; \mathrm{GHz}$ and its on-resonance coupling strength with the qubit to be $g \approx 2 \pi \times 28.9 \; \mathrm{MHz}$ for a quartz BAW resonator. This coupling strength is much larger than typical qubit and mechanical resonator decay rates \cite{Chu2017, Renninger2018}, allowing the coupled system to operate in the regime of strong coupling and therefore efficient swap operations to be performed.

We can now model the idealised coupled dynamics In the rotating frame the standard Jaynes-Cummings Hamiltonian with qubit coupling strengths $g_{l,m}$ and frequencies of the discrete phonon modes $\omega_{l,m}$. Therefore, the Hamiltonian describing the coupled system is 
\begin{ceqn}
\begin{equation} \label{Hamilcoup1}
    \begin{split}
    H=&\delta_{q} \sigma_{+} \sigma_{-} + \sum_{l,m} \delta_{l, m} b_{l, m}^{\dagger} b_{l, m} \\ &+ g_{l, m}\left(\sigma_{+} b_{l, m}+\sigma_{-} b_{l, m}^{\dagger}\right).
    \end{split}
\end{equation}
\end{ceqn}
Here $\delta_q$ and $\delta_{l,m}$ is the detuning from the frequency $\omega_{l,0}$ of the qubit and $l,m$ resonator mode respectively, $g_{l,m}$ is the coupling strength to the $l,m$ resonator mode, $\sigma_{+}$ and $\sigma_{-}$ are the raising and lowering operators for the qubit and $b_{l, m}^{\dagger},$ $b_{l, m}$ are the creation and annihilation operators respectively for the $l,m$ resonator mode.

It is sufficient to only consider the $m = 0$ to $m = 3$ transverse modes and adjacent longitudinal modes ($l = 39, m = 0$) and ($l = 41, m = 0$) in eq.~\eqref{Hamilcoup1} since the effect of the coupling to the other modes is negligible.

In the experimentally relevant regime where the mechanical resonator has a high Q-factor (${\omega_{l,m}\gg\gamma_r, \gamma_q}$), the dynamics of the system are described with the phenomenological Linblad master equation \cite{militello2007}
\begin{ceqn}
\begin{equation} \label{fullcoupledmaster}
    \begin{split}
    \dot{\varrho}&=\frac{-i}{\hbar}[H, \varrho] +\gamma_{q} \mathcal{D}\left[\sigma_{-}\right]\varrho \\ &+(\Gamma_{QP} + \Gamma_P )\mathcal{D}\left[\sigma_{+}\right] \varrho+\frac{\gamma_{\varphi}}{2} \mathcal{D}\left[\sigma_{z}\right]\varrho \\&+ \sum_{l,m} \left[\gamma_r(1 + \bar{n}_{\mathrm{th}}) \mathcal{D}[b_{l,m}] \varrho + (\gamma_r\bar{n}_{\mathrm{th}} + \dot{n}_c)\mathcal{D}[b^{\dag}_{l,m}]\varrho  \right],
    \end{split}
\end{equation}
\end{ceqn}
which accounts for all known sources of decoherence. $H =  \sum_{l, m} g_{l, m}\left(\sigma_{+} b_{l, m}+\sigma_{-} b_{l, m}^{\dagger}\right)$ is the standard Jaynes-Cummings Hamiltonian (see App.~\ref{coupleappendix}), $\varrho$ is the density matrix, $\bar{n}_{\mathrm{th}} = \left(e^{\hbar \omega_{l,m} / k_{B} T}-1\right)^{-1}$ is the mechanical mean thermal occupancy (where $k_B$ is the Boltzmann constant and $T$ is the temperature of the resonator), $\mathcal{D}$ is the dissipating superoperator $\mathcal{D}[L] \hat{\rho}=L \hat{\rho} L^{\dagger}-\frac{1}{2}\left(L^{\dagger} L \hat{\rho}+\hat{\rho} L^{\dagger} A\right)$. The subscripts $_{l,m}$ refer to the $l, m$ mechanical modes, which are for simplicity, assumed to have equal dissipation rates rates $\gamma_r$. $\Gamma_{QP}/2\pi = 30 \; \mathrm{Hz}$ and $\Gamma_P/2\pi = 0.5 \; \mathrm{Hz}$ are the heating rates in the qubit due to quasiparticle poisoning and the Purcell effect respectively, while $\gamma_{\varphi}/2\pi = 0.3 \; \mathrm{MHz}$ is the qubit dephasing rate \cite{Boissonneault2009} and $\gamma_q/2\pi = 27 \; \mathrm{kHz}$ is the qubit dissipation rate \cite{Chu2017}.


In order to solve this master equation, we assume a qubit decay rate of $\gamma_q/2\pi \approx 27 \; \mathrm{kHz}$, a mechanical decay rate of $\gamma_r/2\pi \approx 1 \; \mathrm{Hz}$ and a dephasing rate of $\gamma_{\varphi}/2\pi = 0.3 \; \mathrm{MHz}$, consistent with existing experiments \cite{Boissonneault2009}. We assume the joint system is initially in the state $\ket{g}\ket{n_{l = 503, m = 0} = 1}$, where $\ket{n_{l , m} = 1}$ denotes the $l, m$ mode in the $n = 1$ Fock state and all other modes in the ground state. Using our previously calculated coupling strength of $g \approx 2 \pi \times 28.9 \; \mathrm{MHz}$, we find the system has an efficiency of $\eta_{\mathrm{swap}} \approx 0.8$ after a single swap operation, with evidence of minor off-resonance coupling to adjacent modes.

\begin{figure}[h] 
\begin{center}
\includegraphics[scale  = 0.6]{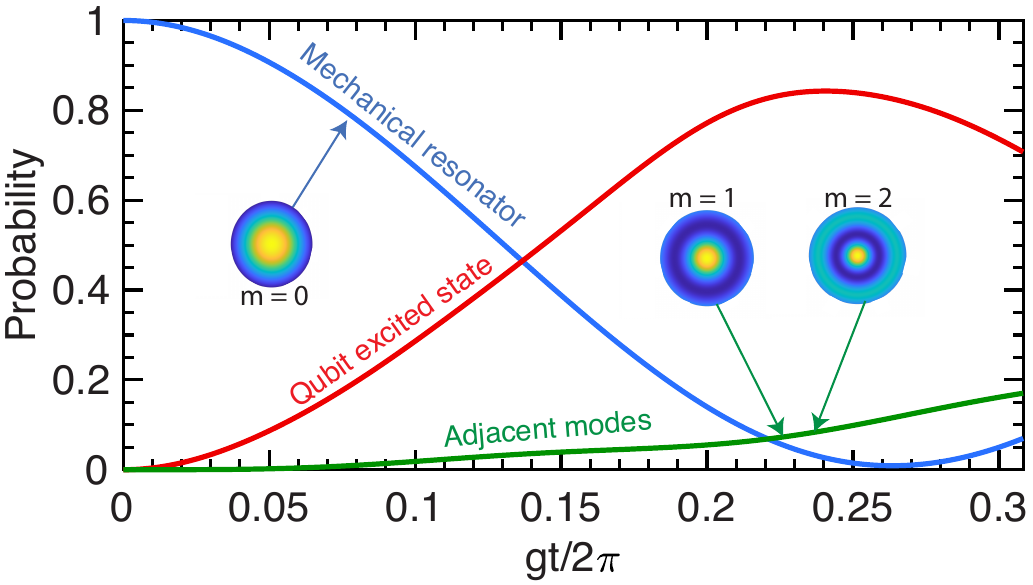}
    \caption{\label{efficiencyplot} Occupancy probabilities for the qubit (red), BAW resonator $l = 40$, $m = 0$ mode (blue), and off-resonant BAW modes (green). The x-axis is normalised to the coupling strength $g$. }
\end{center}
\end{figure}

\section{Gaussian measurement distribution and state discrimination error} \label{appendixsde}
During dispersive readout, quantum noise on the microwave field and dispersive measurement inefficiency leads to a Gaussian statistical distribution for the measured phase for each qubit state, which is known as \textit{state discrimination error}. The centre of the Gaussian for the ground and excited state is $x_0$ and $x_1$ respectively with each having the same standard deviation $\sigma$. The discrimination point $\left(x_d = x_0 + \frac{(x_1 - x_0)}{a}\right)$ is the threshold phase that separates measurements of the qubit state as $\ket{e}$ from those of $\ket{g}$. Typically $a = 2$ is chosen such that the probability of a false measurement of the qubit in the $\ket{e}$ state (false positives) is equal to the probability of a false measurement of the qubit in the $\ket{g}$ state (false negatives) \cite{Sank2014}. In this case the state discrimination error is $\epsilon_{\text {sde }}=\frac{1}{2} \operatorname{erfc}\left[\frac{\left|x_{0}-x_{1}\right|}{2 \sqrt{2 \sigma^{2}}}\right]$ and we can identify  the signal to noise ratio as $\mathrm{SNR}~=~\frac{\left( x_0 - x_1 \right)^2}{2 \sigma^2}$ such that $\epsilon_{\text {sde }}$ can be used as a measure of the $\mathrm{SNR}$. As mentioned in the main text, we require false positives to be suppressed as much as possible. This can be done by shifting the discrimination point, such that $a < 2$.

\begin{figure}[h] 
\begin{center}
\includegraphics[scale  = 0.58]{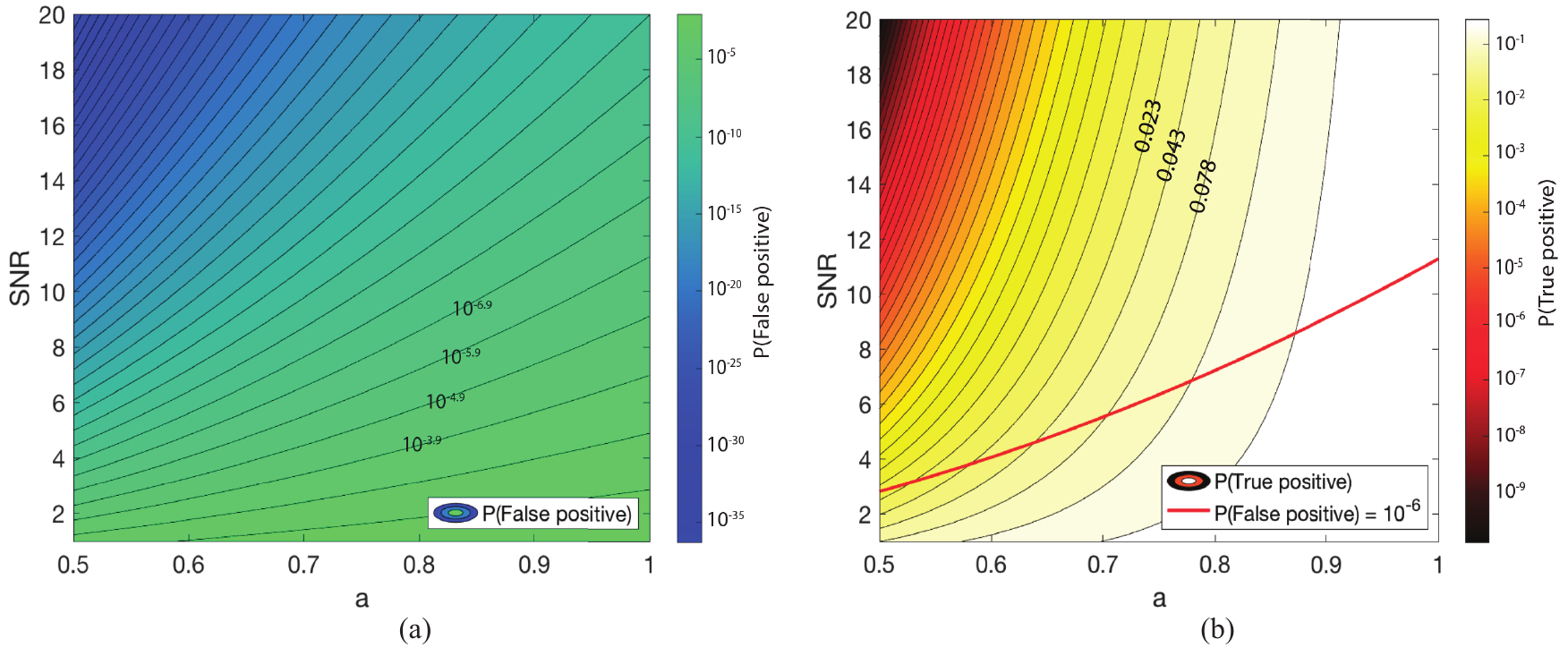}
    \caption[State discrimination error]{\label{SDEcontour} (a) Contour plot of the probability of a false positive for the parameter space of $\mathrm{SNR}$ and $a$. (b) Contour plot of the probability of successfully distinguishing a true positive for the parameter space of $\mathrm{SNR}$ and $a$. The red line displayed in the parameter space corresponds to the values for $a$ and $\mathrm{SNR}$ for which the false positive probability is suppressed to $10^{-6}$. Crucially there is a region of the $\mathrm{SNR}$ and $a$ parameter space where the probability of a false positive is suppressed below $P_{\mathrm{CSL}} \approx 10^{-6}$ while the probability of a true positive still remains sufficiently large  $P( \mathrm{True} \; \ket{1}) \sim 1$.}
\end{center}
\end{figure}

Shifting the discrimination point from the traditional choice of $a = 2$ leads to the following modified expressions for the probability of detecting a false positive and true positive:

\begin{ceqn}
\begin{equation} \label{esepfalse}
    \begin{split}
    P( \mathrm{False} \; \ket{e}) &= \frac{1}{2} \operatorname{erfc}\left[ \frac{\sqrt{\mathrm{SNR}}}{a}\right] \\
    P( \mathrm{True} \; \ket{e}) &= \frac{1}{2} \operatorname{erfc}\left[ \frac{\sqrt{\mathrm{SNR}}(1 - a)}{a}\right].
    \end{split}
\end{equation}
\end{ceqn}
In order to distinguish collapse-induced heating above state discrimination error, we need to suppress $P( \mathrm{False} \; \ket{e})$ while maintaining a reasonably high $P( \mathrm{True} \; \ket{e})$. Fig.~\ref{SDEcontour} displays compares $P( \mathrm{False} \; \ket{e})$ and $P( \mathrm{True} \; \ket{e})$ for a large range of the applicable $(a, \mathrm{SNR})$ parameter space. We find that there is a regime in the parameter space of $(a, \mathrm{SNR})$ for which it is possible to exponentially suppress the error in false positives, while maintaining a reasonable $P( \mathrm{True} \; \ket{e})$. For $a \geq 0.8$ and $\mathrm{SNR} \geq 10$ we find that $P( \mathrm{False} \; \ket{e})$ can be suppressed below $P_{\mathrm{CSL}} \approx 10^{-6}$, while maintaining $P( \mathrm{True} \; \ket{e}) \gtrapprox 0.16$. 

Ref.~\cite{Walter2017} reports the demonstration of rapid ($\sim 50 \; \mathrm{ns} $) high SNR single shot dispersive readout of a transmon qubit. Using similar experimental parameters to those we assumed in the main text, a $\tau = 56 \; \mathrm{ns}$ measurement of the qubit state resulted in a state discrimination error of only $\epsilon_{\mathrm{SDE}} \approx 0.002$. This corresponds to a relatively high signal to noise ratio of $\mathrm{SNR} \approx 17$. Therefore, rapid single shot dispersive readout that sufficiently suppresses $ P( \mathrm{False} \; \ket{e})$ below the collapse-induced occupation number ($P_{\mathrm{CSL}} \approx 2 \times 10^{-7}$) is feasible with state-of-the-art quantum technology.

\section{Heating due to quasiparticles and the Purcell effect} \label{QPappendix}
Recent measurements found a residual excited state population due to quasiparticle excitations in a transmon qubit of $P(\ket{e}) \approx 3 \times 10^{-3}$ \cite{KulikovAnatoly2020METo}. From their qubit decay rate of $\gamma_q/2\pi \approx 100 \; \mathrm{kHz}$, we estimate the quasiparticle induced excitation rate to be $\Gamma_{\mathrm{QP}}/2\pi \approx 300 \; \mathrm{Hz}$. As discussed in the main text, we expect current technology to mitigate quasiparticles should be able to reduce this to $\Gamma_{\mathrm{QP}}/2\pi \approx 30 \; \mathrm{Hz}$.

The Purcell effect is the enhancement of the qubit's spontaneous excitation rate due to the coupling of the qubit to the microwave resonator which is in turn coupled to a transmission line \cite{Sete2014}. The coupling to the transmission line produces spontaneous excitation of the qubit at the rate \cite{Sete2014}

\begin{ceqn}
\begin{equation}
    \Gamma_{P} \approx \frac{\kappa g^{2}}{16 \Delta^{2}}\left(\frac{\bar{n}}{n_{\mathrm{crit}}}\right)^{2}. 
\end{equation}
\end{ceqn}
 For the experimental parameters we have proposed in previous sections, we estimate the Purcell rate to be $\Gamma_P/2 \pi \approx 26 \; \mathrm{Hz} $. A Purcell filter can be employed to reduce this rate by a further factor of $\sim 50$ to $\Gamma_P/2 \pi \approx 0.5 \; \mathrm{Hz}$ \cite{Sete2015}. A swap operation will switch the spurious excited state population of the qubit into the resonator and switch the CSL-induced occupation number of the resonator into the qubit excited state population. 

Here we solve the master equation displayed in Eq.~\eqref{fullcoupledmaster} with the addition of the term $\left(\Gamma_{Q P}+\Gamma_{P}\right) \mathcal{D}\left[\sigma_{+}\right] \hat{\rho}$ to account for quasiparticle and Purcell induced heating of the qubit.

Consider the qubit to be in equilibrium with quasiparticle poisoning induced excitation rate $\Gamma_{\mathrm{QP}}$, decay rate $\gamma_{\mathrm{\mathrm{QP}}}$, Purcell induced heating rate $\Gamma_P$, decay rate $\gamma_P$ and thermally induced excitation rate $\dot{n}_{\mathrm{th},q}$ and decay rate $\gamma_q$. In this case, the equilibrium qubit occupancy $P(\ket{e})$, relates the sum of the excitation and decay rates according to
\begin{ceqn}
\begin{equation} \label{dbalance}
    \Gamma_{\mathrm{QP}} + \Gamma_{P} + \dot{n}_{\mathrm{th},q} = P(\ket{e})\left( \gamma_{\mathrm{QP}} + \gamma_{\mathrm{P}} + \gamma_{q} \ \right).
\end{equation}
\end{ceqn}
As discuss in the main text, $\dot{n}_{\mathrm{th},q}$ is suppressed at milli-Kelvin temperatures. Furthermore the dominant source of decay is $\gamma_{q}$, especially with the addition of a quasiparticle trap and Purcell filter to suppress $\gamma_{\mathrm{QP}}$ and $\gamma_{\mathrm{P}}$. Therefore, $\dot{n}_{\mathrm{th},q}$, $\gamma_{\mathrm{QP}}$ and $\gamma_{\mathrm{P}}$ can be neglected, and using $(\Gamma_{\mathrm{QP}} + \Gamma_P)/2 \pi \approx 30 \; \mathrm{Hz} $ and again assuming a qubit decay rate of $\gamma_q/2 \pi = 27 \; \mathrm{kHz}$, the spurious excited state population of the qubit is
\begin{ceqn}
\begin{equation}
    P(\ket{e}) \approx \frac{\Gamma_{\mathrm{QP}} + \Gamma_P}{ \gamma_q} \approx 1 \times 10^{-3}. 
\end{equation}
\end{ceqn}
Therefore, the initial state of the joint BAW resonator-qubit coupled system is 
\begin{ceqn}
\begin{equation} \label{initqp2}
    \begin{split}
    \hat{\rho}(0) &\approx 1 \times 10^{-3}\ket{e}\ket{\mathrm{vac}}\bra{e}\bra{\mathrm{vac}} \\
    &+ (1 - 1 \times 10^{-3} )\ket{g}\ket{\mathrm{vac}}\bra{g}\bra{\mathrm{vac}}.
\end{split}
\end{equation}
\end{ceqn}
Here, we have assumed that there are is no occupancy of the $n = 1$ Fock state in the resonator due to CSL, since we are interested in the spurious noise rather than the CSL signal. Solving the master equation for this initial state with a qubit-BAW resonator coupling strength of $g/2\pi \approx 28.9 \; \mathrm{MHz}$, we find that the qubit excited state population is only suppressed to $ P(\ket{e}) \sim 10^{-5}$ due to dephasing noise. However if the qubit is tuned to another mode on a timescale short compared to the period of a Rabi oscillation and another swap operation is performed, the spurious heating can be further suppressed. In other words, the qubit can be cooled by swapping spurious heating due to quasiparticle poisoning and the Purcell effect into the mechanical resonator. The saturation limit to how significantly the qubit can be cooled using this protocol is the ratio of the spurious excitation rate and the qubit-resonator coupling strength ~$\frac{\Gamma_{\mathrm{QP}} + \Gamma_P}{g_{l,m}}$. The initial population of the qubit is $\frac{\Gamma_{\mathrm{QP}} + \Gamma_P}{\gamma_q}$, therefore our protocol is effective in cooling the qubit in the strong-coupling regime $g_{l,m} > \gamma_q$. We predict this relation through a detailed balance equation similar to Eq.~\eqref{dbalance} in which coupling to the mechanical resonator is modelled as providing a decay channel for the qubit. After each swap operation the qubit is tuned on-resonance with another mechanical mode and another swap operation is performed in order to cool the qubit further, until the saturation limit is obtained.

We now numerically confirm the predicted saturation limit of how much the qubit can be cooled using our proposed procedure. If no heating additional to the noise level corresponding to the collapse rate $\lambda_{\mathrm{min}}$ is observed in the proposed experiment, then any value for the collapse rate of $\lambda_c \geq \lambda_{\mathrm{min}}$ is strictly excluded. Therefore, we will firstly solve the master equation for the initial state 
\begin{ceqn}
\begin{equation} \label{initqp}
    \begin{split}
    \hat{\rho}(0) &\approx 7 \times 10^{-7}\ket{e}\ket{\mathrm{vac}}\bra{e}\bra{\mathrm{vac}} \\
    &+ (1 - 7 \times 10^{-7} )\ket{g}\ket{\mathrm{vac}}\bra{g}\bra{\mathrm{vac}}.
\end{split}
\end{equation}
\end{ceqn}
which assumes the dominant source of heating in the qubit is quasiparticle poisoning. We will then compare to the solution of the master equation with an initial state that assumes no quasiparticle or Purcell-induced heating $\Gamma_{\mathrm{QP}} = \Gamma_P = 0$, such that the only source of heating in the coupled system is CSL-induced heating:
\begin{ceqn}
\begin{equation}
    \begin{split}
    \hat{\rho}(0) &\approx 5.5\times 10^{-7}\ket{g}\ket{l,m}\bra{g}\bra{l,m} \\
    &+ (1 - 5.5 \times 10^{-7} )\ket{g}\ket{\mathrm{vac}}\bra{g}\bra{\mathrm{vac}}.
\end{split}
\end{equation}
\end{ceqn}
In this way, we compare the expected excited state population due to CSL alone to the excited state population due to quasiparticle poisoning and the Purcell effect. Assuming Bassi \textit{et. al}'s lowest estimate for the collapse rate lower bound of $\lambda_c = 10^{-12} \; \mathrm{s}^{-1}$, the probability a higher-order mechanical mode $l \sim 40$ will be in the $n = 1$ Fock state due to collapse-induced heating is $P \approx 5.5 \times 10^{-7}$. This probability scales linearly with the collapse rate $\lambda_c$. Therefore, we identify the minimum testable collapse rate as the collapse rate for which the excited state probability due to CSL intersects the excited state probability due to noise sources after a single swap operation. The quasiparticle-induced heating is suppressed to its minimum possible value after a single qubit-mechanical resonator swap operation. The absence of any qubit excited state population above $P(\ket{e}) \approx 7 \times 10^{-7}$ would conclusively falsify CSL for collapse rates $\lambda_c \geq 1.5 \times 10^{-12} \; \mathrm{s}^{-1}$ at the expected correlation length ($r_c \approx 100 \; \mathrm{nm}$).

\section{Measurement Induced heating} \label{measinducedapp}
Non-QND effects arise due to higher order terms in $g/\Delta$ in the dispersive approximation. Most significantly, the drive Hamiltonian acts as creation and annihilation operators on eigenstates of the dispersive Hamiltonian, while the eigenstates of the system in the lab frame are instead eigenstates of the full interaction Hamiltonian $H = \hbar \omega_{r} a^{\dagger} a-\hbar \omega_{q} \frac{\sigma_{z}}{2}+\hbar g\left(a^{\dagger} \sigma_{-}+a \sigma_{+}\right)$. Mismatch between these dressed eigenstates and the bare eigenstates of the dispersive Hamiltonian causes the drive Hamiltonian ($H_d$) to induce qubit transitions in the dressed basis: 
\begin{ceqn}
\begin{equation}
  H_d = \varepsilon^{*} a+\varepsilon a^{\dagger} \approx\left(\epsilon^{*} \bar{a}+\epsilon \bar{a}^{\dagger}\right)-\frac{g}{\Delta}\left(\epsilon^{*} \bar{\sigma}_{-}+\epsilon \bar{\sigma}_{+}\right),
\end{equation}
\end{ceqn}
where $a$ and $\bar{a}$ are the bare and dressed creation operators respectively. This mismatch between the bare and dressed creation and annihilation operators in the readout Hamiltonian generates qubit excited state population in the dressed basis given by $P_{H} \approx \left|\frac{\epsilon g}{\Delta^2 }\right|^2$ \cite{Khezri2016}. For a measurement drive of $\epsilon/2\pi = 25 \; \mathrm{MHz}$ (allowing the readout resonator to be populated with 2.5 photons within the first 10 $\mathrm{ns}$ being driven), a coupling strength and detuning of $g/2\pi =  100 \; \mathrm{MHz}$ and $\Delta/2\pi = 3 \; \mathrm{GHz}$ respectively we obtain  $P_{H} \approx 8 \times 10^{-8}.$ 

\section{Thermal Noise}
A large thermal noise background has precluded previous experiments from resolving collapse-induced heating within the expected CSL parameter range \cite{Vin2017, Vin2020}. This is because these experiments have operated with low frequencies in the high temperature limit ($k_B T \gg \hbar\omega$), requiring them to resolve CSL-induced heating above this thermal noise background. This necessitates a precise knowledge of the thermal noise level, and can lead to inconsistent interpretations \cite{Vin2017}. Here this problem is solved by accessing the regime where $k_B T \ll \hbar\omega$~\cite{Forstner2020}. Taking $\omega = \omega_q = \omega_r$ and $k_B T \ll \hbar\omega$, the thermal heating rates are $\dot{n}_{\mathrm{th,r}} \approx \frac{\gamma_r}{2 \pi}\exp\left(- \frac{\hbar \omega}{k_B T} \right)$ and $\dot{n}_{\mathrm{th,q}} \approx \frac{\gamma_q}{2 \pi}\exp\left(- \frac{\hbar \omega}{k_B T} \right)$ in the mechanical resonator and the qubit respectively.  An upper bound on the minimum testable collapse rate limited by thermal noise is therefore $\lambda_{c, \mathrm{min}} = (\dot{n}_{\mathrm{th},r} + \dot{n}_{\mathrm{th},q})/D$. Assuming the system is thermalised to the base-plate of a dilution fridge at $T = 10 \; \mathrm{mK}$, while taking $\omega/2\pi = 6.33 \; \mathrm{Hz}$, we obtain a minimum testable collapse rate of $\lambda_{c, \mathrm{min}} \sim 10^{-20} \; \mathrm{s}^{-1}$, many orders of magnitude below both Adler and Bassi's predicted lower bounds.

\section{Coloured CSL} \label{colappendix}
One of the weaknesses of the white noise CSL model is that it is incompatible with the collapse noise field having a physical origin \cite{CarlessoMatteo2018Ccmf}. If the collapse noise field has a cosmological origin, it should have a non-white spectrum with a high frequency cut-off \cite{Adlerr2007}. Therefore, it is important to consider the implications of a non-white spectrum for the bounds on the CSL model's phenomenological parameters $\lambda_c$ and $r_c$. In order to examine how a non-white spectrum influences the decoherenece rate we examine the formulation of CSL as a random potential \cite{CarlessoMatteo2018Ccmf}:
\begin{ceqn}
\begin{equation}
    \hat{V}_{\mathrm{cCsL}}(t)=-\frac{\hbar \sqrt{\lambda}}{m_{0}} \int \mathrm{d} \mathbf{z} \hat{M}(\mathbf{z}) w(\mathbf{z}, t).
\end{equation}
\end{ceqn}
Here, $w(\mathbf{z}, t)$ is a classical Gaussian noise characterised by 
\begin{ceqn}
\begin{equation}
    \mathbb{E}[w(\mathbf{z}, t_1)]=0, \quad \mathbb{E}[w(\mathbf{z}, t_1) w(\mathbf{x}, t_2)]=\delta^{(3)}(\mathbf{z}-\mathbf{x}) f(t_1-t_2),
\end{equation}
\end{ceqn}
where $\mathbb{E}[\cdot]$ denotes the stochastic average. For the correlation function $f(t)=\delta(t)$ the simple white-noise CSL model is obtained. In order to model the effects of a coloured CSL noise spectrum, we consider an exponentially decaying noise correlation function:
\begin{ceqn}
\begin{equation}
    f(t_1-t_2)=\frac{\Omega_{c}}{2} e^{-\Omega_{c}|t_1-t_2|}.
\end{equation}
\end{ceqn}
Here, $\Omega_{c}$ is the high frequency cut-off, and for $\Omega_{c} \rightarrow \infty$ we obtain the delta function correlation function corresponding to the simple white-noise CSL model. For an oscillator of resonance frequency $\omega$, the coloured CSL noise spectrum modifies the CSL-induced noise spectral density for the oscillator to be \cite{CarlessoMatteo2018Ccmf}
\begin{ceqn}
\begin{equation}
    S_{\mathrm{CSL}} = S_{\mathrm{c}} \cdot \frac{\Omega_{\mathrm{c}^2}}{\Omega_{\mathrm{c}}^2 + \Omega^2 },
\end{equation}
\end{ceqn}
while in the simple white noise CSL model, the CSL-induced heating rate is $\dot{n}_c = \lambda_c D$ (where $D$ is the CSL cross-section of the oscillator). In the coloured CSL model the CSL-induced heating rate is
\begin{ceqn}
\begin{equation}
    \dot{n}_c = \lambda_c D \frac{\Omega_{\mathrm{c}^2}}{\Omega_{\mathrm{c}}^2 + \Omega^2 }.
\end{equation}
\end{ceqn}
If the collapse noise field has a cosmological origin, it is expected to be at $\sim 10^{10} - 10^{11} \; \mathrm{Hz}$. In this case, we note that for $\Omega_{\mathrm{c}} = 10^{10} \; \mathrm{Hz}$ and $\Omega_{\mathrm{c}} = 10^{11} \; \mathrm{Hz}$ there is only a minor reduction from the white noise CSL-induced heating rate of $30 \%$ and $0.4 \%$ respectively (assuming a mechanical mode frequency of $\omega/2\pi = 6.33 \; \mathrm{GHz}$). Therefore, our proposed experiment has the capability to test coloured CSL models.

\section*{References}

\footnotesize \renewcommand{\refname}{\vspace*{-30pt}}  
\bibliographystyle{apsrev4-2}

\bibliography{bib_Supp}